\documentclass[%
preprint,
 amsmath,amssymb,
 aps,
 pre,
]{revtex4-1}
\usepackage{graphicx}% Include figure files
\usepackage{dcolumn}% Align table columns on decimal point
\usepackage{bm}
\usepackage[utf8]{inputenc}   % inutile avec LuaLaTeX / XeLaTeX
\usepackage[T1]{fontenc}
\usepackage{mathtools}
\usepackage{titlesec}
\usepackage{footnote}
\usepackage{amsmath}
\usepackage{amssymb}
\usepackage{caption}
\usepackage{subcaption} % Pour les sous-figures
\usepackage{epstopdf}
\usepackage{enumerate}
\usepackage{siunitx} % Pour avoir les unités SI
\usepackage{hyperref} % Pour les liens hypertext
\usepackage{cleveref} % Pour les jolies références
\usepackage{threeparttable} % Pour les notes dans les tableaux
\usepackage{makecell}
\usepackage{enumitem}
\usepackage{accents}
\usepackage{float}
\usepackage[table]{xcolor}
\usepackage{setspace}
\usepackage{algorithm} % Package pour l'environnement algorithm
\usepackage{algpseudocode}

\usepackage[section]{placeins}
\newcommand{\aver}[1]{\langle {#1} \rangle}

\newcommand{\RRR}{{\mathsf  R}}
\newcommand{\BBB}{{\mathsf  B}}
\newcommand{\SSS}{{\mathsf  S}}
\newcommand{\DDD}{{\mathsf  D}}
\newcommand{\RBSD}{{\mathsf  R,\mathsf B,\mathsf S,\mathsf D}}
\newcommand{\RBSDtext}{$\mathsf{R}$, $\mathsf{B}$, $\mathsf{S}$ and $\mathsf{D}$}
\newcommand{\II}{{\mathsf  I}}

\newcommand{\pa}[1]{\overline{#1}}

\newcommand{\CC}{{\pa{C}}}
\newcommand{\cNN}{{\overline{C}}}
\newcommand{\kNN}{k}
\newcommand{\eNN}{\varepsilon}
\newcommand{\omegNN}{\omega}
\newcommand{\vNN}{b}

\newcommand{\LNN}{L}
\newcommand{\KNN}{K}
\newcommand{\ENN}{E}
\newcommand{\VNN}{B}

\newcommand{\Cmu}{\mathcal{C}_\mu}
\newcommand{\sigmac}{\sigma_c}
\newcommand{\sigmaK}{\sigma_k}
\newcommand{\sigmaeps}{\sigma_\varepsilon}
\newcommand{\sigmaV}{\sigma_b}
\newcommand{\Cepsz}{{\mathcal{C}_{\varepsilon 0}}}
\newcommand{\Cepsd}{\mathcal{C}_{\varepsilon 2}}
\newcommand{\Cchi}{\mathcal{C}_{\chi}}

\newcommand{\KepsS}{$k$--$\varepsilon$--$b$ }
\newcommand{\pred}[1]{\hat{#1}}

\newcommand{\xx}{{\bf  x}}
\newcommand{\uu}{{\bf  u}}

\newcommand{\UU}{{\bf  U}}

\newcommand{\dd}[2]{\frac{\mathrm{d}#1}{\mathrm{d}#2}}
\newcommand{\ie}{\textit{i.e. }}
\newcommand{\softplus}{\mathrm{Softplus}}
\newcommand{\expnumber}[2]{{#1}\mathrm{e}{#2}}

\newcommand{\Fr}{\mathrm{Fr}}
\newcommand{\KdK}{\tfrac{\kNN_d}{\kNN}}

\newcommand{\fococ}{{f_\cNN}}
\newcommand{\focok}{{f_\kNN}}
\newcommand{\focoe}{{f_\eNN}}
\newcommand{\focov}{{f_\vNN}}
\newcommand{\DFr}{\Delta \Fr}
\newcommand{\DKdK}{\Delta \KdK}

\begin{document}

\preprint{APS/123-QED}

\title{Learning Turbulence Closures with Physics-Informed Neural Networks for the Rayleigh--Taylor Transition to Turbulence}% Force line breaks with \\

\author{Paul Creusy$^{1,2}$, Beno\^it-Joseph Gr\'ea$^{1,2}$, Antoine Briard$^1$ and Teo Granger$^{1,2}$}
\affiliation{%
    $^1$CEA, DAM, DIF, F-91297 Arpajon, France \\
    $^2$Universit\'e Paris-Saclay, CEA, LMCE, F-91680 Bruy\`eres-le-Ch\^atel, France \\
}%

\date{\today}% It is always \today, today,

\begin{abstract}
    Reynolds-averaged Navier--Stokes (RANS) turbulence models are known to perform poorly in predicting the dynamics of Rayleigh--Taylor mixing when turbulence is not fully developed, particularly during the transition from an initially perturbed interface. In this work, we investigate the use of data-driven strategies to enhance a simple $k$--$\varepsilon$--$b$ model for this transitional regime.
    The turbulence model is first embedded within a surrogate physics-informed neural network (PINN), enabling the calibration of coefficients that account for parametric errors and the identification of corrective terms representing structural errors associated with missing physical processes.
    The learned corrections are then re-expressed onto the model state variables and relevant flow indicators, leading to explicit analytical modifications of the closure.
    The resulting fully interpretable corrected model is assessed against an extensive database of direct numerical simulations (DNS) of Rayleigh--Taylor flows. This framework enables improved predictions of the mixing-layer growth during the transition to turbulence.
    \begin{description}
        \item[Usage]
        \item[PACS numbers]
        \item[Structure]
    \end{description}
\end{abstract}

\pacs{Valid PACS appear here}% PACS, the Physics and Astronomy
%Use showkeys class option if keyword
\maketitle

%\tableofcontents

\section{Introduction}

The Rayleigh--Taylor instability \cite{Rayleigh_1882,Taylor_1950}, which occurs when a heavy fluid is placed above a lighter one in a downward gravitational field, is one of the most widely studied hydrodynamic instabilities owing to its importance in numerous astrophysical, geophysical, and engineering applications \cite{Zhou2017a,Zhou2017b}. In particular, it plays a critical role in inertial confinement fusion (ICF), where mixing between the ablator and the deuterium--tritium fuel has a major impact on the capsule yield \cite{Lindl_1995,Betti_2016,Remington2019,Zhou_2025}.

The dynamics of the Rayleigh--Taylor instability proceeds through several distinct regimes \cite{Sharp1984}. Starting from an initially perturbed interface, the perturbation amplitude first undergoes exponential growth in the linear regime \cite{Chandrasekhar2013}. This is followed by a nonlinear potential-flow regime, in which bubbles and spikes evolve at an approximately saturated velocity \cite{Goncharov2002}. As secondary instabilities develop, this regime becomes unstable and transitions toward turbulence \cite{Cook2001}. At late times, provided that the influence of external boundaries remains negligible, the flow reaches a self-similar turbulent mixing  regime \cite{Youngs1984}, in which the mixing-layer width $L$ evolves as
\begin{equation}
    L(t) = 2\,\alpha_\infty\,\mathcal{A}\,g\,(t - t_\infty)^2,
\end{equation}
where $g$ is the acceleration due to gravity, $\mathcal{A}$ is the Atwood number characterizing the density contrast between the two fluids, and $t$ denotes time.

Within the Boussinesq approximation, corresponding to the low-Atwood-number limit, the mixing zone is statistically symmetric, as bubbles and spikes evolve in a similar manner. This symmetry leads to a single set of self-similar parameters, where $\alpha_\infty$ is the asymptotic growth-rate coefficient and $t_\infty$ is the virtual origin associated with the self-similar solution.

Notably, it has been shown that while the asymptotic growth rate is nearly universal, with $\alpha_\infty \simeq 0.02$ (the precise value depending on the definition of $L$), the virtual origin $t_\infty$ is highly sensitive to the initial conditions and, in particular, reflects the transition to turbulence \cite{Thevenin2025,Thevenin2025b}.

To model the transition to turbulence, a wide range of approaches have been proposed in the literature, ranging from simple phenomenological descriptions—such as buoyancy--drag formulations \cite{Ramshaw1998,Dimonte2000,Youngs2020,Schilling2020}—to more elaborate models that explicitly account for multimode nonlinear interactions \cite{Haan1989,Haan1991,Ofer1996}.

In practice, these transition models are often used to initialize Reynolds-Averaged Navier--Stokes (RANS) closures \cite{Rollin2013} once the flow is assumed to have entered a fully turbulent regime. However, this procedure is not without difficulties: it is not always possible to unambiguously determine when the flow has become fully turbulent, and one must still consistently assimilate the output of the transition model into the RANS framework.

An alternative approach to modelling Rayleigh--Taylor dynamics consists in embedding the physics of transition directly within the RANS framework. This approach relies on the fact that key quantities controlling transition can be reconstructed or inferred from the model state variables, for instance through suitably chosen indicators~\cite{Xiao2025,Qi2025} or effective initial conditions~\cite{Thevenin2025}. Since this reconstruction problem is inherently ill-posed, it naturally motivates data-driven strategies, in particular those based on direct numerical simulations (DNS).

This strategy has already been explored in other contexts, notably through field inversion and machine learning (FIML) techniques applied to boundary-layer flows~\cite{Duraisamy2019,Duraisamy2021}. Such an approach, however, presupposes the ability to distinguish between different sources of error in the RANS model, in particular parametric errors, associated with model calibration, and structural errors, arising from the functional form of the closures~\cite{Xiao2019}.

Different strategies are commonly used to calibrate RANS models for the prediction of buoyancy-driven flows. The most straightforward approach consists in exploiting self-similar or analytical solutions to constrain the model coefficients. However, this strategy does not guarantee accurate behavior during transient phases. By contrast, data-driven calibration based on a diverse set of flow configurations is generally more robust. In this respect, various techniques can be employed, such as static and dynamic calibration procedures \cite{Boureima2022}, optimization via genetic algorithms \cite{Gimenez2019} or Bayesian approaches \cite{Edeling2014,Nadiga2019}. The latter is particularly attractive, as a large spread in the posterior distribution of the calibrated coefficients is a strong indicator of the presence of structural model errors. More recently, physics-informed neural networks (PINNs) \cite{Raissi2019} have been proposed for the RANS equations \cite{Eivazi2022} and the calibration of RANS models including the Rayleigh--Taylor instability. These approaches offer the advantage of not requiring a dedicated RANS solver, as the RANS model can be incorporated directly into the loss function \cite{Xiao2023,Zhang2026}.

Moreover, it has been demonstrated that, by augmenting PINNs with an additional neural network while enforcing the model equations in the loss function, parametric and structural errors can be disentangled, thereby enabling the detection of missing physics in a model~\cite{Zou2024,Patel2024}. Following this strategy, the objective of the present work is to assess whether a relatively simple turbulence model, such as a corrected $k$--$\varepsilon$--$b$ closure~\cite{Kenjeres2000,Schilling2017}, where $k$ denotes the turbulent kinetic energy, $\varepsilon$ its dissipation rate, and $b$ the concentration variance, is able to capture the transition to turbulence in the Rayleigh--Taylor instability.

This work is organized as follows. We first present the configuration and the database of simulations used in this study. The next section is devoted to classical model calibration, as well as calibration using PINNs to disentangle parametric and structural errors. The final section focuses on the assimilation and reinterpretation of structural errors using symbolic regression, with the aim of improving the baseline model to account for the transition to turbulence.

\section{Database and surrogate model}
In this study, dedicated to the transition to turbulence in Rayleigh--Taylor miscible flows, we make use of an extensive database of $484$ DNS \cite{Thevenin2025data}, previously presented in \cite{Thevenin2025,Thevenin2025b}. These simulations were performed at a resolution of $1024^2 \times 2048$ points using our in-house pseudo-spectral code \textsc{Stratospec} \citep{Grea2019,Briard2020,Briard2025,Briard2025b}, which solves here the incompressible Navier--Stokes equations under the Boussinesq approximation, corresponding to the low-Atwood-number $\mathcal A$ limit, for the velocity $\UU(\xx,t)$, together with an advection--diffusion equation for the concentration of heavy fluid $C(\xx,t)$, since the heavy and light fluids are miscible.

In addition, by introducing the characteristic length $(\nu^2/\mathcal{A}g)^{1/3}$ and time $(\nu/(\mathcal{A}g)^2)^{1/3}$, constructed from the kinematic viscosity $\nu$ and the reduced acceleration $\mathcal A g$, the quantities and governing equations can be non-dimensionalized, following \cite{Schilling2017,Thevenin2025}. Assuming further that the kinematic viscosity $\nu$ and the diffusion coefficient $\kappa$ are constant and equal, we obtain
\begin{subequations}
    \begin{align}
        \boldsymbol{\nabla} \boldsymbol{\cdot} \UU                     & = 0, \label{eq:basea}                                                                 \\
        \partial_t \UU + \UU \boldsymbol{\cdot} \boldsymbol{\nabla}\UU & = -\boldsymbol{\nabla} \Pi - 2 C \, \boldsymbol{e}_z + \nabla^2 \UU, \label{eq:baseb} \\
        \partial_t C + \UU \boldsymbol{\cdot}\boldsymbol{\nabla} C     & = \nabla^2 C, \label{eq:basec}
    \end{align}
\end{subequations}
where $\Pi$ is the reduced pressure.

The initial concentration field (or equivalently the density field) is constructed such that the heavy fluid lies above the lighter fluid in a downward gravitational field of magnitude $g$, directed along the $z$ axis. The two fluids are separated by a diffuse interface that is modulated by two-dimensional random perturbations with an annular spectrum, as introduced in \cite{Dimonte2004} and detailed in \cite{Thevenin2025}. In contrast, the initial velocity field is set to zero, $\UU(\xx,0)=0$.

Finally, the simulations are parameterized by four non-dimensional numbers, namely the initial Reynolds number $\mathsf{R}$, defined using the mean perturbation length scale and a reduced vertical acceleration, the mean initial interface steepness $\mathsf{S}$, the renormalized bandwidth of the perturbation $\mathsf{B}$, and the renormalized diffusive thickness $\mathsf{D}$. For convenience, we will write a set of initial conditions as $\mathsf{I}=\{ \mathsf{R},\mathsf{B},\mathsf{S},\mathsf{D} \}$.

We define the Reynolds average as the average over the horizontal $(x,y)$-plane, corresponding to the statistically homogeneous directions of the flow. Quantities averaged in this way are denoted by $\overline{(\cdot)}$. Any flow variable $Q$ can then be decomposed as
\[
    Q (\xx,t) = \overline{Q}(z,t) + q(\xx,t),
\]
where $q$ denotes the fluctuation about the mean.

In the Boussinesq limit, it can be shown that the mean velocity vanishes, i.e. $\overline{\mathbf{U}} = 0$.
The mean flow is therefore fully characterized by the mean concentration field, which satisfies the transport equation,
\begin{equation}
    \frac{\partial \CC}{\partial t}=-\frac{\partial (\overline{u_z c})}{\partial z}+ \frac{\partial^2 \CC}{\partial z^2}. \label{eq:C}
\end{equation}
Consequently, the evolution of the mean concentration field is governed not only by molecular diffusion, but also by the turbulent concentration flux $\overline{u_z c}$, which needs to be determined by a turbulent model.

The mixing-layer width $L(t)$ can be defined from the mean concentration field.
Among several possible definitions, we adopt the classical integral definition \citep{Andrews1990}
\begin{equation}
    L = 6 \int_{-\infty}^{+\infty} \overline{C}\,(1-\overline{C})\, dz,
\end{equation}
where the prefactor $6$ ensures that $L$ coincides with the layer thickness for a piecewise linear mean concentration profile.

Reynolds-averaged quantities, hereafter referred to as ``1D'' quantities since they depend only on the vertical coordinate $z$, can be further reduced to ``0D'' quantities by integration along the vertical direction and normalization by the mixing-layer width. Accordingly, from the turbulent kinetic energy density $k=\overline{\uu\cdot\uu}/2$, its viscous dissipation rate $\varepsilon$, and the concentration fluctuation variance $b=\overline{cc}$, we define the corresponding 0D kinetic energy $K$, dissipation $E$, and concentration variance $B$ as
\begin{equation}
    K=\frac{1}{L} \int_{-\infty}^{+\infty} k \, dz, \qquad
    E=\frac{1}{L} \int_{-\infty}^{+\infty} \varepsilon \, dz, \qquad
    B=\frac{1}{L} \int_{-\infty}^{+\infty} b \, dz,
\end{equation}
which provide useful global diagnostics of the Rayleigh--Taylor mixing layer.

\begin{figure}
    \centering
    \includegraphics[width=1 \linewidth]{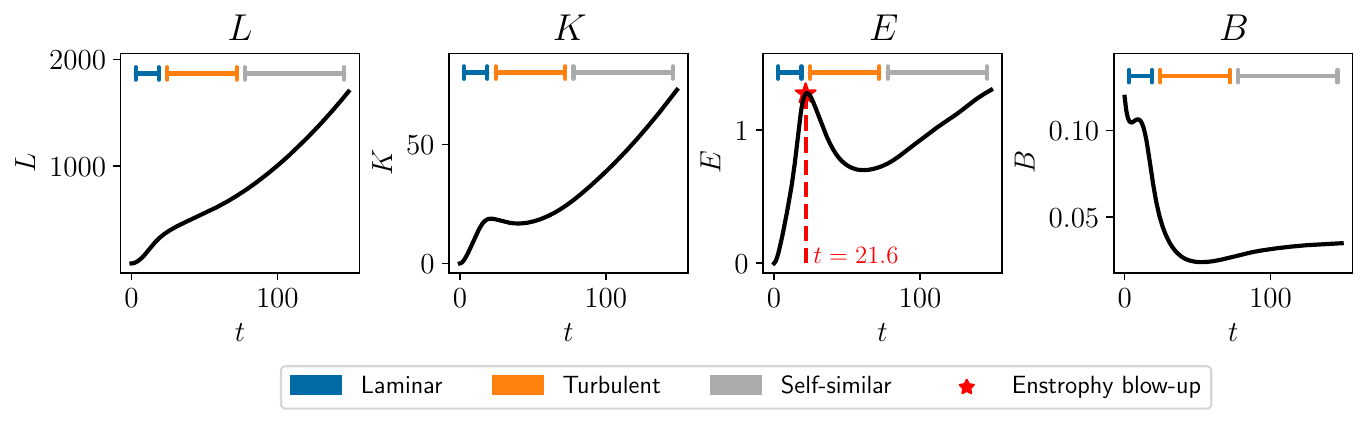}
    \caption{Evolution of volume-averaged quantities (0D) in time from DNS defined by $\RRR = 25.2$, $\BBB = 1.23$, $\SSS = 3$ and $\DDD=3$. Four distinct regime appear, first a linear one in which perturbations grow exponentially followed by non-linear regime while the flow is still laminar. After the enstrophy blow-up, the mixing zone transitions to turbulence at its center and reaches after some time a self-similar regime.}
    \label{fig:RT_phases}
\end{figure}

\begin{figure}
    \begin{center}
        \includegraphics[width=\textwidth]{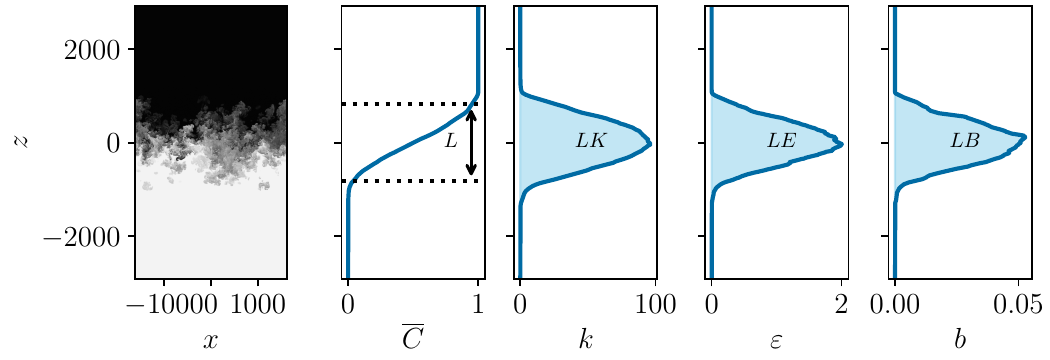}
        \caption{Two-dimensional snapshot of the heavy-fluid mass fraction field, together with one-dimensional profiles of the mean concentration $\overline{C}$, turbulent kinetic energy $k$, dissipation rate $\varepsilon$ and scalar variance $b$ in the turbulent regime, extracted from the DNS database described in \cite{Thevenin2025data} ($\RRR = 25.2$, $\BBB = 1.23$, $\SSS = 3$, $\DDD=3$) at $t=146.1$ on a high resolution $2048 \times 2048 \times 4096$ mesh. \label{fig:setup}}
    \end{center}
\end{figure}

In Fig.~\ref{fig:RT_phases}, we illustrate the temporal evolution of the 0D quantities for a representative DNS from the database, highlighting the different regimes encountered during the growth of the mixing layer. In Fig.~\ref{fig:setup}, we show a typical two-dimensional snapshot from the same simulation in the turbulent regime, together with the corresponding one-dimensional profiles and their zero-dimensional integrals.

Although the DNS database covers a large portion of the initial $\RBSD$ domain, it is convenient to encapsulate its results by developing a surrogate model for the 0D and 1D quantities. We use physics-informed neural networks (PINNs)~\cite{Raissi2019} to construct such a surrogate, taking the initial conditions and time as inputs and predicting the corresponding 0D and 1D quantities. One major advantage of imposing physics-informed constraints is that the surrogate gains mild extrapolation capabilities, both in time and in the initial-condition space $\mathsf{I}$.

These neural networks incorporate a set of hard and soft constraints in the loss function, including the 0D and 1D evolution equations of the quantities, as described in \cite{Thevenin2025}. This procedure ensures accurate and consistent predictions, even in regions of the parameter space where DNS data are sparse. The surrogate model has been successfully used to infer and assess the sensitivity to initial conditions in \cite{Thevenin2025,Thevenin2025b}. In the following, we exploit this surrogate to calibrate and improve RANS solutions for RT configurations.

\section{Turbulence model and calibration}
\subsection{A classical \texorpdfstring{$k$--$\varepsilon$--$b$}{K-eps-b} turbulence model for the Rayleigh-Taylor instability}
In this section, we present the equations of a simple turbulence model with state variables given by the turbulent kinetic energy density $k(z,t)$, its dissipation rate $\varepsilon(z,t)$, and the concentration variance density $b(z,t)$ \cite{Schilling2017,Hanjalic2002}. These quantities define a turbulent viscosity (non dimensional) as $\nu_t =\mathcal C_\mu k^2/\varepsilon$. The equations for the model are given by

\begin{subequations}
    \begin{equation}
        \frac{\partial k}{\partial t}
        =
        \frac{\partial}{\partial z} \left[\left(1+ \frac{\nu_t}{\sigma_k}\right) \frac{\partial k}{\partial z} \right]
        +2 \frac{\nu_t}{\sigma_c} \frac{\partial \CC}{\partial z}
        -\varepsilon,
        \label{eq:k}
    \end{equation}
    \begin{equation}
        \frac{\partial \varepsilon}{\partial t}
        =
        \frac{\partial}{\partial z} \left[\left(1+ \frac{\nu_t}{\sigma_\varepsilon} \right) \frac{\partial \varepsilon}{\partial z} \right]
        +2 \mathcal C_{\varepsilon 0} \frac{\varepsilon}{k}\frac{\nu_t}{\sigma_c} \frac{\partial \CC}{\partial z}
        -\mathcal C_{\varepsilon 2}\frac{\varepsilon^2}{k},
        \label{eq:e}
    \end{equation}
    \begin{equation}
        \frac{\partial b}{\partial t}
        =
        \frac{\partial}{\partial z} \left[ \left(1+\frac{\nu_t}{\sigma_b}\right) \frac{\partial b}{\partial z} \right]
        +2 \frac{\nu_t}{\sigma_c} \left(\frac{\partial \CC}{\partial z} \right)^2
        -2 \mathcal C_\chi \frac{\varepsilon}{k} b,
        \label{eq:s}
    \end{equation}
\end{subequations}

In the system of Eqs.~\eqref{eq:k}--\eqref{eq:s}, the terms on the right-hand side represent, respectively, physical and turbulent diffusion, buoyancy production, and dissipation.

We do not introduce an additional equation for the dissipation of the concentration variance. Instead, this term is closed algebraically using the turbulent frequency $\varepsilon/k$ and the coefficient $\mathcal C_\chi$, as is commonly proposed, for instance, in \cite{Schilling2017}.

Notably, the turbulent quantities $k$, $\varepsilon$ are coupled to the mean concentration field through Eq.~\eqref{eq:C}, using the first-gradient closure for the vertical concentration flux,
\begin{equation}
    \overline{u_z c} = -\frac{\nu_t}{\sigma_c} \partial_z \overline{C}. \label{eq:uc}
\end{equation}

Although this $k$-$\varepsilon$-$b$ model includes a transport equation for the concentration variance $b$, it is important to note that, within the first-gradient closure framework of Eq.~\eqref{eq:uc}, $b$ does not feed back either on the mean flow or on the other turbulent quantities, namely $k$ and $\varepsilon$. One objective of the present study is precisely to introduce such a dependence through a data-driven strategy.

Therefore, the turbulence model is fully specified by eight coefficients, namely $\sigma_{k,\varepsilon,b,c}$ and $\mathcal C_{\varepsilon 0,\varepsilon 2,\chi, \mu}$, which must be calibrated.

\subsection{Classical calibration}

\subsubsection{Methodology}

The calibration of a RANS model can first be carried out on solutions of simplified turbulent flows, foremost among them the decay of homogeneous isotropic turbulence. This makes it possible to set the value of $\mathcal C_{\varepsilon 2}$ to 1.92, as is commonly done in most RANS models \cite{Meldi2012}. Similarly we fix the coefficient setting the turbulent viscosity at $\mathcal C_\mu=0.09$ \cite{Schiestel2010}.

When the calibration is performed for the Rayleigh--Taylor instability, it is often carried out in the late-time self-similar regime, in particular to reproduce the self-similar growth rate $\alpha _\infty$. Owing to the availability of a large DNS database, we instead propose in this section to perform the calibration of the remaining coefficients directly on the data, and then to check the consistency of the results with the late-time self-similar regime.

To begin with, we use the 0D system detailed in App.~\ref{sec:0D} to calibrate the model. This approach makes it possible to determine the values of the coefficients related to production and dissipation, but not those related to diffusion, which will be discussed later. Following \cite{Nadiga2019,Boureima2022}, we consider several calibration strategies.

The first approach, referred to as \emph{static}, consists in taking samples of the initial parameters $\II$ from the surrogate of the DNS database and performing a linear regression, as detailed in Sec.~\ref{sec:regression}. The regression is carried out in the late-time turbulent regime, using times chosen either in the interval $t \in [80,150]$ or in the interval $t \in [100,150]$ for a stricter late-time calibration. Note that the transition to turbulence, identified from the local maximum of enstrophy, occurs around $t=30$, although the precise value depends on the initial conditions.

The second approach, referred to as \emph{dynamic}, is a Bayesian calibration (see for instance \cite{Edeling2014}) performed on fully integrated trajectories, also sampled from the surrogate model, with initial conditions taken at $t=80$, as described in App.~\ref{sec:bayesian}.

\begin{figure}[ht!]
    \centering
    \includegraphics[width=1 \linewidth]{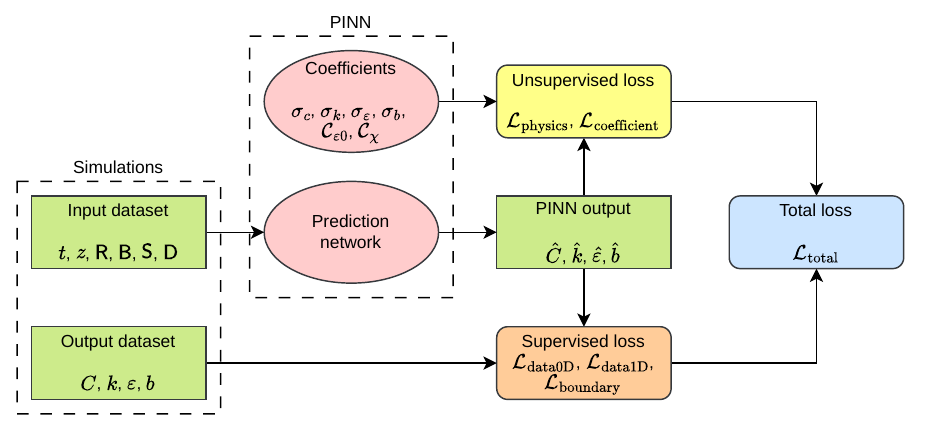}
    \caption{Description of the calibration framework PINN-$\mathcal C$. Arrays of numbers are represented in green squares, neural networks in pink ellipses and loss terms with rounded rectangles. The standard supervised term allows the PINN to learn the data and defines the boundary conditions. The coefficients to calibrate are defined as weights of the PINN and act through the physics-informed term of the loss. The information transits to the prediction network through the supervised loss and then to the coefficients through the unsupervised loss.}
    \label{fig:PINN_diagram}
\end{figure}

The third approach PINN-C, which can also be classified as \emph{dynamic}, relies on a PINN that takes as inputs the initial conditions $\II$ and the time $t$, and returns as outputs the 0D state vector of the turbulence model, namely $L$, $K$, $E$, and $B$. The loss function used for training is decomposed into a supervised part, based on data from the surrogate model starting at $t=80$, and an unsupervised part, which enforces the 0D equations, as illustrated in Fig.~\ref{fig:PINN_diagram} and detailed in App.~\ref{sec:pinn}. In this setup, the model coefficients are therefore learned during the training phase so as to fit both the data and the 0D equations of the $k$-$\varepsilon$-$b$.

We verify that the PINN provides an accurate solution of the 0D equations, since it yields the same solution as that obtained by direct integration of the 0D system. Notably, the initial conditions corresponding to the PINN solution, i.e., the 0D quantities at time $t=80$, are learned during training and may therefore differ slightly from the initial values given by the data. This feature defines a data assimilation procedure for initializing the model.

Similarly and in order to calibrate the full set of coefficients, including those related to diffusion, we also propose an extended calibration approach PINN-C, which directly provides the 1D solutions of the RANS model and enforces the full system of equations, namely Eq.~\eqref{eq:C} and Eqs.~\eqref{eq:k}--\eqref{eq:s}.

\subsubsection{Results}

The calibration results obtained on the Rayleigh--Taylor dataset, following the procedures detailed above, are reported in Table~\ref{tab:coef}. As can be seen, the calibrated coefficients obtained from the different approaches---regression, Bayesian calibration, and PINN-C 0D or 1D---are in good agreement. In particular, one first observes that calibrating the model over the ranges $t \geq 80$ or $t \geq 100$ (for the regression) leads to similar results, indicating that the calibration procedure is well converged.

\begin{table}[ht!]
    \begin{ruledtabular}
        \begin{tabular}{lcccccccc}
            \textbf{Source}                & $\sigmac$                       & $\sigmaK$ & $\sigmaeps$ & $\sigmaV$ & $\Cepsz$                        & $\Cepsd$  & $\Cchi$                         \\
            \colrule
            Standard \cite{Schilling2017}  & 0.6-1.48                        & 0.87-1.0  & 1.3         & -         & 0.815-0.95                      & 1.90-1.92 & 1.5                             \\
            Schilling \cite{Schilling2017} & 0.08                            & 0.09      & 0.10        & 0.07      & 1.44                            & 2.32      & 0.51                            \\
            Lin. Reg.                      & 0.143                           & -         & -           & -         & 1.361                           & 1.92      & 0.864                           \\
            Lin. Reg. late                 & 0.144                           & -         & -           & -         & 1.341                           & 1.92      & 0.912                           \\
            Bayesian                       & $0.135 \pm \expnumber{5.5}{-5}$ & -         & -           & -         & $1.375 \pm \expnumber{2.2}{-4}$ & 1.92      & $0.926 \pm \expnumber{5.5}{-4}$ \\
            PINN-C 0D                      & 0.142                           & -         & -           & -         & 1.319                           & 1.92      & 0.985                           \\
            PINN-C 1D                      & 0.137                           & 0.111     & 0.188       & 0.119     & 1.345                           & 1.92      & 0.922
        \end{tabular}
    \end{ruledtabular}
    \caption{Comparison of the \KepsS coefficient values from different sources and methods. All of them use $\Cmu=0.09$. The $\sigmaK$, $\sigmaeps$ and $\sigmaV$ columns are left empty for the methods that rely solely on volume-averaged data as they do not appear in the equations. The Bayesian method intervals are given at 95\% (mean values are rounded to the third decimal, making rounding dominant compared to the error interval, the error interval should therefore be taken as a mesure of standard deviation). Posterior distributions are available in Fig.~\ref{fig:mcmc_posterior}.}
    \label{tab:coef}
\end{table}

\begin{figure}[ht!]
    \centering
    \includegraphics[width=1 \linewidth]{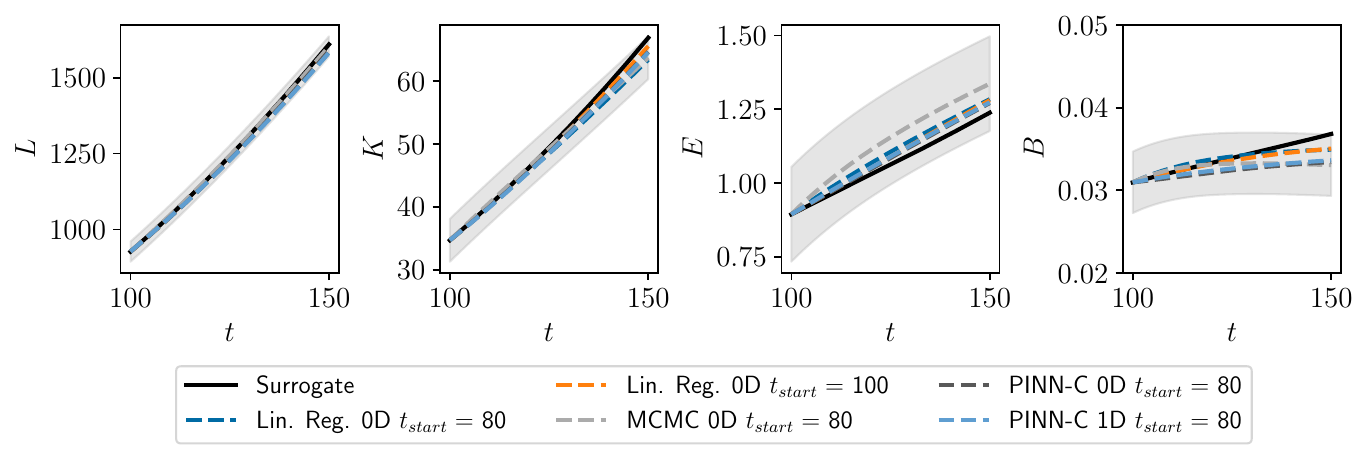}
    \caption{Comparison between the trajectories obtained by integration of the calibrated equations and the surrogate model on a simulation defined by $\RRR = 14.8$, $\BBB = 0.06$, $\SSS = 5.35$ and $\DDD=1.46$. The transparent gray area represents the 95\% error margins of the MCMC method.}
    \label{fig:calibration_0D_curves}
\end{figure}

\begin{table}[ht!]
    \begin{ruledtabular}
        \begin{tabular}{ccccccccc}
            Calibration interval & $\sigma_c$ & $\sigma_k$ & $\sigma_\varepsilon$ & $\sigma_b$ & $\mathcal{C}_{\varepsilon 0}$ & $\mathcal{C}_{\varepsilon 2}$ & $\mathcal{C}_{\chi}$ & $\alpha_M$ Eq.~\eqref{eq:alpha_from_coefs} \\
            \colrule
            $t \in [1, 150]$     & 0.133      & 0.146      & 0.245                & 0.265      & 1.411                         & 1.92                          & 0.958                & 0.0142                                     \\
            $t \in [15,150]$     & 0.129      & 0.159      & 0.288                & 0.336      & 1.421                         & 1.92                          & 0.881                & 0.0138                                     \\
            $t \in [30,150]$     & 0.128      & 0.1        & 0.18                 & 0.124      & 1.365                         & 1.92                          & 0.879                & 0.0188                                     \\
            $t \in [50,150]$     & 0.135      & 0.108      & 0.186                & 0.122      & 1.354                         & 1.92                          & 0.913                & 0.0188                                     \\
            $t \in [80,150]$     & 0.137      & 0.111      & 0.188                & 0.119      & 1.345                         & 1.92                          & 0.922                & 0.0195                                     \\
        \end{tabular}
    \end{ruledtabular}
    \caption{Comparison of the \KepsS coefficient values when changing the calibration interval for the PINN-C 1D method. The upper bound of the time interval remains unchanged at 150. The values vary significantly for early calibration showing a high underlying structural error.\label{tab:coef1}}
\end{table}

The model coefficients determine the asymptotic self-similar solution. From the 0D equations, the self-similar growth rate $\alpha_M$ of the model can be estimated as \cite{Grea2015}
\begin{equation}
    \label{eq:alpha_from_coefs}
    \alpha_M=\frac{\mathcal C_\mu}{\sigma_c}\frac{(\mathcal C_{\varepsilon 2}-\mathcal C_{\varepsilon 0})^2}{(4 \mathcal C_{\varepsilon 0}- 3)(4 \mathcal C_{\varepsilon 2}- 3)}.
\end{equation}

This predicted value is close to the mean value obtained from the database, see Tab.~\ref{tab:coef1}, which is approximately $0.02$ when the calibration is performed using data for $t \geq 80$. This agreement further supports the legitimacy of the data-based calibration, since it recovers the value obtained from DNS, as detailed in~\cite{Thevenin2025b}.

We now present in Fig.~\ref{fig:calibration_0D_curves} comparisons of the 0D quantities between the DNS surrogate model for a trajectory that was not used during calibration (test trajectory) and the integrated solutions of the $k$--$\varepsilon$--$b$ model. The models are calibrated using the different strategies and initialized with the values of the surrogate solution at $t=100$.

Note that, at this late time, the effect of data assimilation in the PINN becomes negligible, which explains why the model can be initialized directly with the initial conditions from the DNS surrogate.

Therefore, in the fully developed regime, the $k$--$\varepsilon$--$b$ model calibrated on data provides very good predictions of the turbulent quantities. This is in particular confirmed by the model error estimated from the Bayesian calibration, as detailed in App.~\ref{sec:bayesian}, and also shown in Fig.~\ref{fig:calibration_0D_curves}. Notably, the posterior distributions of the calibrated coefficients are tightly clustered around their mean values, as also shown in App.~\ref{sec:bayesian}.

We now turn to the calibration results obtained with the PINN-C 1D, which in particular provide estimates of the diffusion coefficients. The calibrated values of $\sigma_{k,\varepsilon,b}$ are relatively small compared to standard values, of the order of $0.1$--$0.2$, reflecting the enhanced diffusion that characterizes the Rayleigh--Taylor instability. This feature was already highlighted in \cite{Schilling2017}, where the closures were optimized directly against DNS data.

\begin{figure}[ht!]
    \centering
    \includegraphics[width=1 \linewidth]{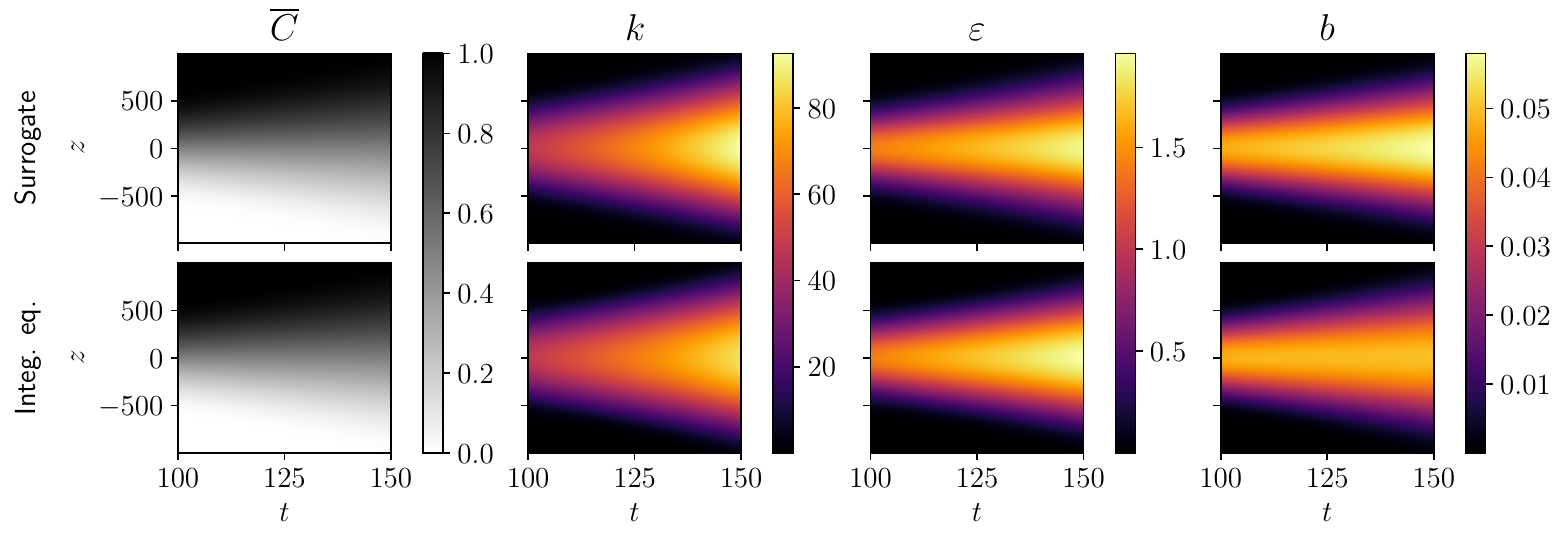}
    \caption{Comparison of horizontal concatenation of 1D vertical profiles in time between the surrogate and the calibrated integrated equations on the test simulation. The calibration is performed using the PINN-C on both 0D and 1D data.    }
    \label{fig:pinn1Dmap}
\end{figure}

In Fig.~\ref{fig:pinn1Dmap}, we compare the integrated solutions of the $k$--$\varepsilon$--$b$ model calibrated with the PINN-C against the 1D profiles of the state variables, again for a trajectory from the DNS surrogate that was not used during training. This again leads to very satisfactory predictions.

\begin{figure}[ht!]
    \centering
    \includegraphics[width=1 \linewidth]{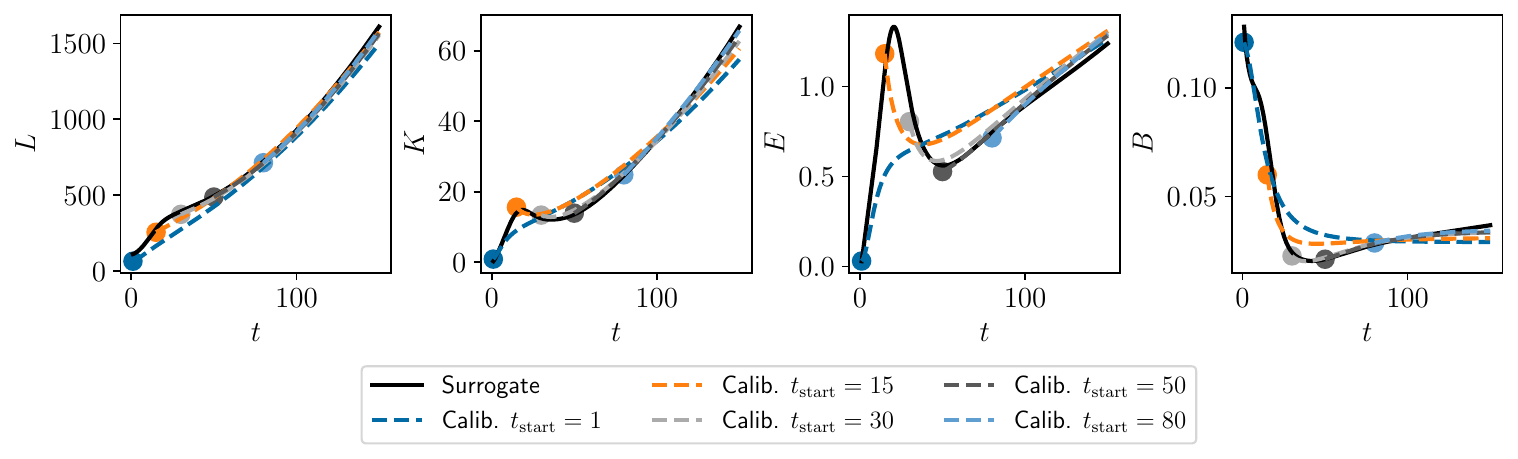}
    \caption{Comparison between the trajectories obtained by integration of the calibrated equations with PINN-C 1D and the surrogate model for several calibration time intervals on a test simulation ($\RRR = 14.8$, $\BBB = 0.06$, $\SSS = 5.35$, $\DDD=1.46$). The $t_\mathrm{start}$ corresponds both to the initial time of the integration and to the lower limit of the time interval used for the calibration. The coefficients values are given in table \ref{tab:coef1}. A calibration starting at $t_\mathrm{start}=1$ ignores both transition features such as the dissipation peak and the self-similar growth rate. At late times, the calibration retrieves the self-similar growth rate.}
    \label{fig:fulltraj}
\end{figure}

Of course, the good results obtained with the $k$--$\varepsilon$--$b$ model calibrated on data in the turbulent regime are to be expected, since this model is specifically designed to describe that regime. By contrast, when the same calibration strategy is applied from the onset of the Rayleigh--Taylor instability, using the PINN-C 1D for instance, the calibrated model performs less well. In Fig.~\ref{fig:fulltraj}, we present the integrated solutions of the $k$--$\varepsilon$--$b$ model calibrated by the PINN-C 1D using data intervals starting from progressively earlier times. Clearly, the accuracy of the model deteriorates during the transition phase around $t=30$ and, more specifically, reveals difficulties in capturing the enstrophy blow-up.

This issue is therefore reflected in the values of the calibrated coefficients, which account not only for parametric errors but also for the structural error associated with the absence of corrections for the transition to turbulence. In Tab.~\ref{tab:coef1}, we observe that including data from the transition regime progressively deteriorates the quality of the solution in the self-similar regime, as $\alpha_M$ no longer recovers the expected value. This deterioration is mainly due to the change in the coefficient $\mathcal C_{\varepsilon 0}$, which is related to the production of dissipation $\varepsilon$.

In the next section, we propose a strategy to disentangle these two components of the error, thereby paving the way for improved closure models during the Rayleigh--Taylor transition to turbulence.

\subsection{Disentangling parametric and structural errors \label{sec:param-struc}}

Therefore, we adopt the same strategy as that proposed by~\cite{Zou2024}, here applied to our RANS $k$--$\varepsilon$--$b$ model. This strategy consists not only in constructing a PINN to reproduce the model solution and calibrate its coefficients against data, as described above, but also in augmenting the model equations with corrective terms represented by a neural network. These corrective terms are intended to capture the model error associated with the transition to turbulence. One advantage of this strategy, compared for instance with a Bayesian approach, is that it does not presuppose a specific form for the model error, such as a Gaussian distribution.

\begin{figure}[ht!]
    \centering
    \includegraphics[width=1 \linewidth]{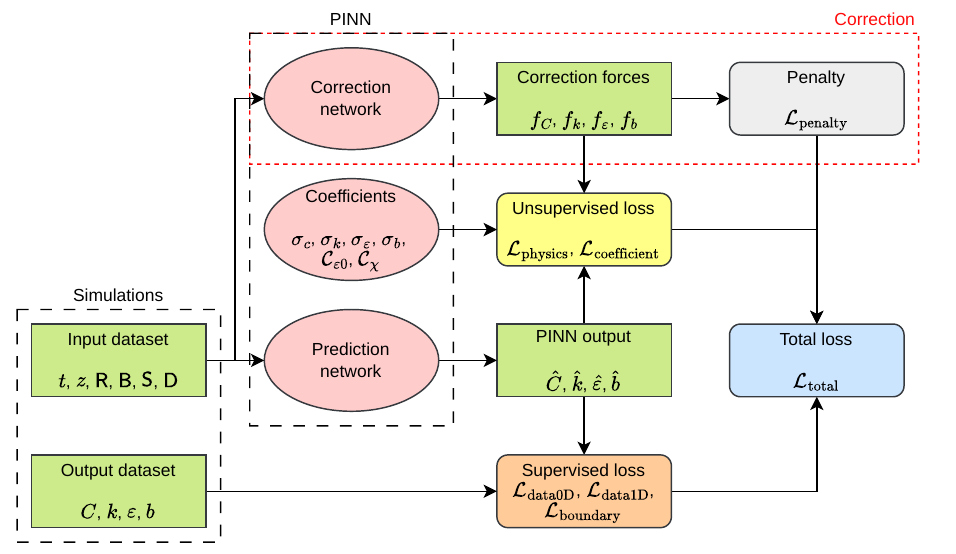}
    \caption{Calibration and correction PINN-C-$f$ framework. The supplementary correction part compared to the previous calibration framework is framed in red dotted line. The correction network takes the same inputs as the prediction one and outputs correction forces acting through the physics loss. A penalty loss term is added to control the weight of the correction in the final result and avoid disrupting the calibration.}
    \label{fig:PINN_diagram2}
\end{figure}

A schematic representation of this PINN-C-$f$ strategy is presented in Fig.~\ref{fig:PINN_diagram2} and detailed in Appendix~\ref{sec:pinn}. Naturally, the additional corrective terms introduced into the model equations must be penalized in the total loss of the PINN; otherwise, the calibration effort would be absorbed entirely by these corrective terms rather than by the model itself. This introduces an additional hyperparameter, $\lambda_{\mathrm{penalty}}$, which controls the penalization of the corrective terms relative to the supervised and unsupervised contributions to the loss.

\begin{table}[ht!]
    \begin{ruledtabular}
        \begin{tabular}{ccccccccc}
            $\lambda_{\mathrm{penalty}}$ & $\sigma_c$ & $\sigma_k$ & $\sigma_\varepsilon$ & $\sigma_b$ & $\mathcal{C}_{\varepsilon 0}$ & $\mathcal{C}_{\varepsilon 2}$ & $\mathcal{C}_{\chi}$ & $\alpha_M$ \\
            \colrule
            $10^{2}$                     & 0.125      & 0.088      & 0.159                & 0.188      & 1.383                         & 1.92                          & 0.925                & 0.0175     \\
            $10^{1}$                     & 0.125      & 0.089      & 0.168                & 0.186      & 1.386                         & 1.92                          & 0.92                 & 0.0172     \\
            $10^{0}$                     & 0.125      & 0.098      & 0.178                & 0.176      & 1.378                         & 1.92                          & 0.937                & 0.018      \\
            $10^{-1}$                    & 0.126      & 0.106      & 0.194                & 0.177      & 1.371                         & 1.92                          & 0.943                & 0.0185     \\
            $10^{-2}$                    & 0.129      & 0.109      & 0.203                & 0.189      & 1.353                         & 1.92                          & 0.923                & 0.0198     \\
            $10^{-3}$                    & 0.131      & 0.112      & 0.201                & 0.158      & 1.344                         & 1.92                          & 0.878                & 0.0205     \\
            $10^{-4}$                    & 0.133      & 0.116      & 0.195                & 0.16       & 1.348                         & 1.92                          & 0.841                & 0.0198     \\
            $10^{-5}$                    & 0.166      & 0.122      & 0.197                & 0.148      & 1.635                         & 1.92                          & 0.609                & 0.0027     \\
        \end{tabular}
    \end{ruledtabular}
    \caption{Values of the coefficients obtained for a calibration with correction forces for various levels of penalization. With a penalization around $10^{-4}$, one retrieves similar values to the ones obtained with a late-time calibration.\label{tab:penalty}}
\end{table}

In Tab.~\ref{tab:penalty}, we show the effect of the hyperparameter $\lambda_{\mathrm{penalty}}$ on the calibration obtained with the PINN-C-$f$ when all DNS data from $t=1$ onward are included. When the corrective terms are strongly penalized, i.e. for large values of $\lambda_{\mathrm{penalty}}$, the calibration results are unsatisfactory (the $\alpha _M$ values are too small), as the structural error is transferred almost entirely to the coefficients, thereby deteriorating the late-time self-similar regime. By contrast, when the corrective terms are insufficiently penalized, the discrepancy is absorbed almost entirely by these terms, as discussed above.

To identify an appropriate compromise, we monitor the value of the model self-similar growth rate $\alpha_M$, which recovers the DNS value for $\lambda_{\mathrm{penalty}} \approx 10^{-3}$. It is worth noting that, for this value of the hyperparameter, the calibrated coefficients are also close to those obtained when calibrating only in the fully turbulent regime ($t \geq 80$) in Tab.~\ref{tab:coef1}. This indicates that the model error has been successfully transferred to the corrective terms, which therefore represent the missing physics associated with the transition regime.

\begin{figure}[ht!]
    \centering
    \includegraphics[width=1 \linewidth]{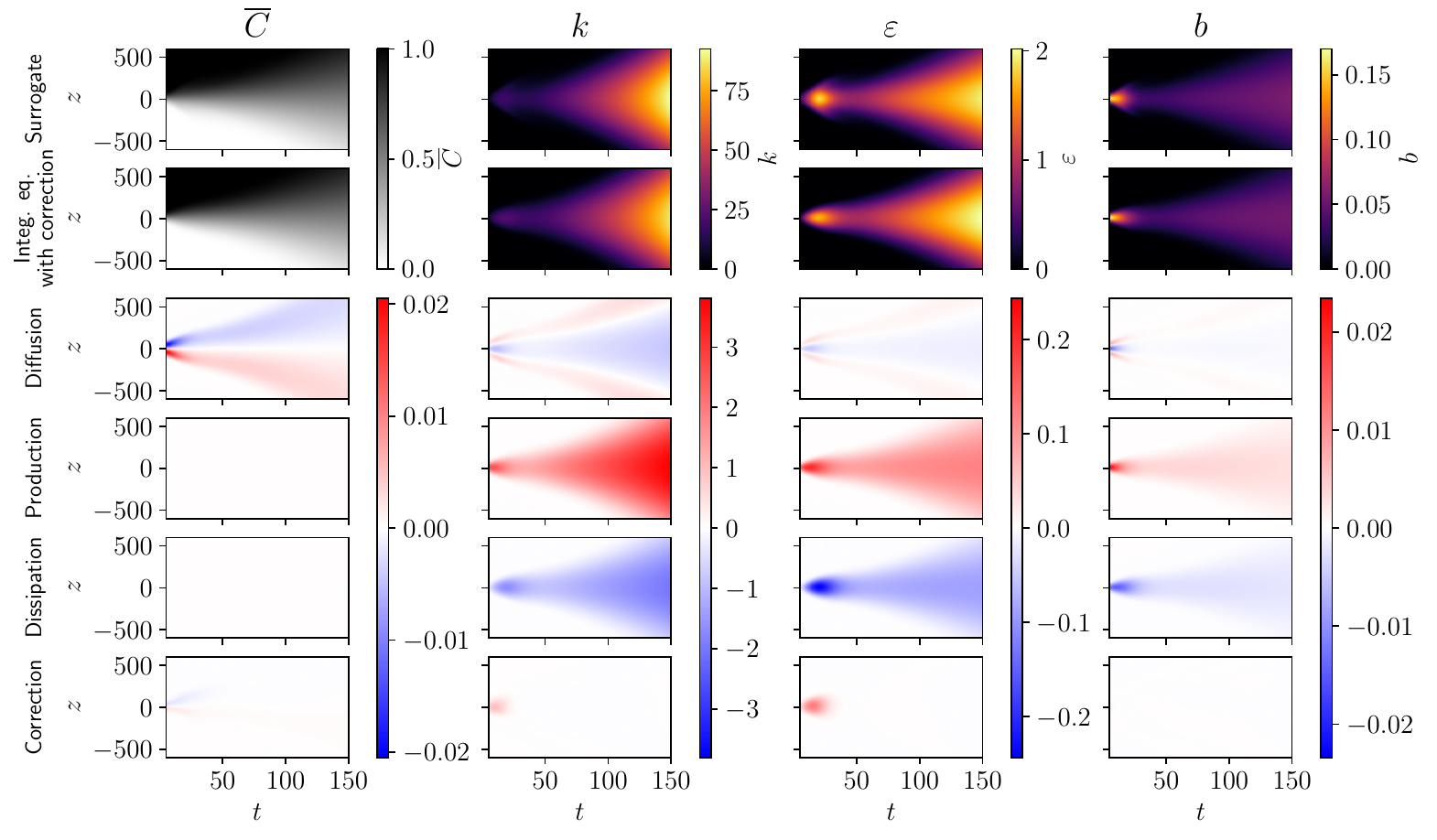}
    \caption{Comparison between the surrogate and an integration of the equations including the correction forces learned by the PINN-C-$f$. The decomposition of all four terms of the equations (diffusion, production, dissipation and correction) is presented for each quantity of the model. The dissipation is counted negatively so that the vertical sum of the four contribution graphs yields the time derivative of the quantity. The correction forces have a notable contribution only in the transition regime and mostly on $\kNN$ and $\eNN$.}
    \label{fig:FoCoPINN_comparison}
\end{figure}

In Fig.~\ref{fig:FoCoPINN_comparison}, we present a comparison of the 1D profiles, for a trajectory not used during training, obtained with the optimized model including corrective terms from PINN-C-$f$. The model is thus able to capture the transition phase efficiently, and in particular the enstrophy blow-up, which was not possible with the baseline model. One may also note that the corrective terms, also shown in the figure, are active only during this transition phase, confirming that this strategy successfully disentangles parametric and structural errors in the RANS model.

The corrected model is, however, not usable in practice, since the corrective terms depend on the initial condition parameters $\II$, as well as on position and time, rather than on the state variables of the model. Addressing this limitation is the objective of the next section.

\begin{figure}[ht!]
    \centering
    \includegraphics[width=1 \linewidth]{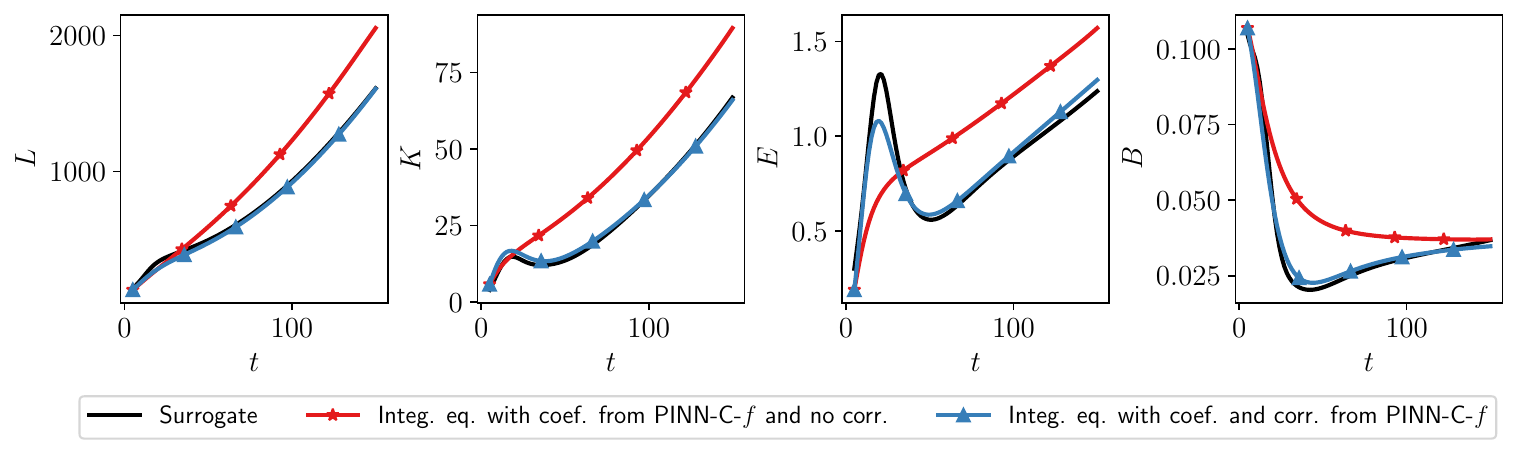}
    \caption{Comparison between the surrogate and two integrations for the usual test simulation ($\RRR = 14.8$, $\BBB = 0.06$, $\SSS = 5.35$, $\DDD=1.46$). The first integrations in red line with stars is performed with the baseline equations of the model and the coefficients calibrated by the PINN-C-$f$. The self-similar growth rate is correct but the transition regime is not. The second integration in blue line with triangles includes the correction forces $f$ predicted with the correction network trained inside the PINN-C-$f$. This reconstruction is more accurate and includes complex features such as the dissipation peak on $\eNN$.\label{fig:foco_integ}}
\end{figure}

\section{Model correction}

The difficulty of the baseline $k$-$\varepsilon$-$b$ model in capturing transition to turbulence arises principally from its inability to reproduce the dissipation peak associated with enstrophy amplification during transition. This suggests that the required model correction should primarily act through the dissipation equation, and possibly through the coefficient $\mathcal{C}_{\varepsilon 0}$.

To assess this hypothesis, we repeat the PINN calibration procedure using the corrective neural network described in Sec.~\ref{sec:param-struc}, with the correction introduced only as a forcing term in the dissipation equation. The results of the PINN-C-$f_\varepsilon$ calibration are reported in Tab.~\ref{tab:definitive_coefs}, showing coefficient calibrations similar to those obtained with the PINN correction applied to all equations. Moreover, Tab.~\ref{tab:companalytical} shows that restricting the correction to the dissipation equation does not deteriorate drastically the model performance compared to the full correction on all the equations.
\begin{table}[ht!]
    \begin{ruledtabular}
        \begin{tabular}{ccccccccc}
            $\sigma_c$ & $\sigma_k$ & $\sigma_\varepsilon$ & $\sigma_b$ & $\mathcal{C}_{\varepsilon 0}$ & $\mathcal{C}_{\varepsilon 2}$ & $\mathcal{C}_{\chi}$ & $\alpha$ \\
            \colrule
            0.126      & 0.106      & 0.195                & 0.187      & 1.358                         & 1.92                          & 0.934                & 0.0198   \\
        \end{tabular}
    \end{ruledtabular}
    \caption{Coefficients values obtained for the PINN-C-$\focoe$ (with a force correction only on $\eNN$) and a penalty of $\lambda_\mathrm{penalty}=10^{-3}$. The analytical model also relies on this set of coefficients.}
    \label{tab:definitive_coefs}
\end{table}

\begin{table}[ht!]
    \begin{ruledtabular}
        \begin{tabular}{ccc}
            Model                  & MRE     & MRE L   \\
            \colrule
            PINN-C                 & 27.12\% & 18.50\% \\
            PINN-C-$f$             & 3.13\%  & 2.18\%  \\
            PINN-C-$f_\varepsilon$ & 4.11\%  & 2.28\%  \\
            PINN-NC-$\Cepsz$       & 5.53\%  & 3.30\%  \\
            Analytical             & 10.18\% & 6.35\%  \\
        \end{tabular}
    \end{ruledtabular}
    \caption{Comparison of the Mean Relative Error (MRE) on the test set obtained with different correction models on the time interval $t \in [40,150]$. The analytical model being based on the PINN correction for $\Cepsz$, the error of this last model can be considered as the lowest achievable for the analytical. The better the symbolic regression, the closer it will be from the PINN. The mean relative error is computed on 100 regularly spaced points on the volume averaged quantities. The dataset is restricted to inertial trajectories with $\mathsf{R} \geq 14$ and $\mathsf{D}\leq 2$.}
    \label{tab:companalytical}
\end{table}

\begin{table}[ht!]
    \begin{ruledtabular}
        \begin{tabular}{lcc}
            Method           & Calibration  & Correction                                                                     \\
            \colrule
            PINN-C           & $\checkmark$ & $\times$                                                                       \\
            PINN-C-$f$       & $\checkmark$ & $t,z,\mathsf{I} \rightarrow \fococ,\focok,\focoe,\focov$                       \\
            PINN-C-$\focoe$  & $\checkmark$ & $t,z,\mathsf{I} \rightarrow \focoe$                                            \\
            PINN-NC-$\Cepsz$ & $\times$     & $\partial_z \hat{\cNN}, \hat{\kNN}, \hat{\eNN}, \hat{\vNN} \rightarrow \Cepsz$ \\
            Analytical       & $\times$     & $\Fr, \KdK \rightarrow \Cepsz$                                                 \\
        \end{tabular}
    \end{ruledtabular}
    \caption{Summary of the various PINN-based strategies for calibration and model correction aimed at capturing the Rayleigh–Taylor transition to turbulence.}
    \label{tab:PINN_details}
\end{table}

\begin{figure}[ht!]
    \centering
    \includegraphics[width=1 \linewidth]{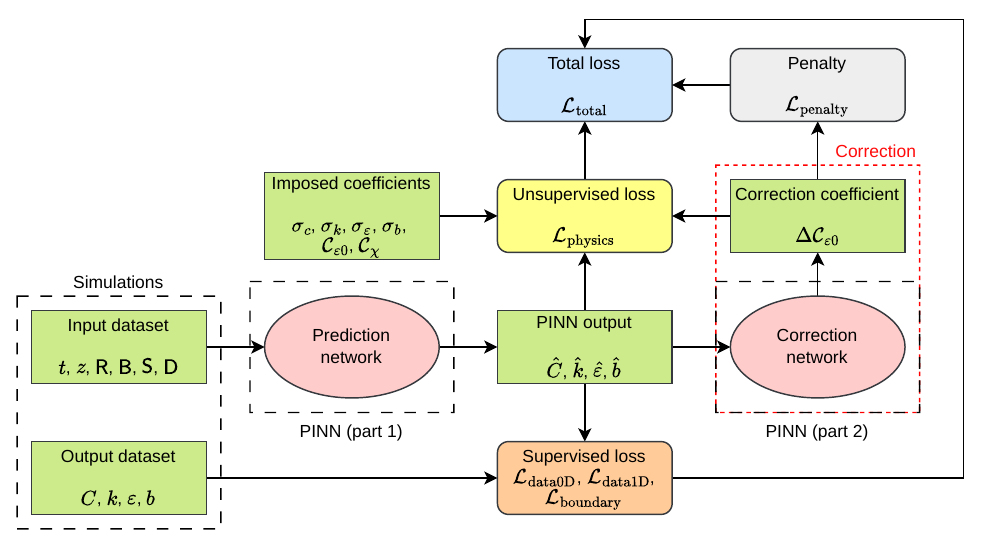}
    \caption{Calibration and correction framework for the PINN-NC-$\mathcal{C}_{\varepsilon 0}$ model. The correction block is highlighted by the red dotted line. In contrast with previous strategies, the correction network takes as input the output of the prediction network, ensuring consistency with the state variables of the system without requiring a projection step that could introduce an additional error term. The PINN is split into two parts only for visualization purposes and not for implementation. An input-transformation layer, not shown here, provides the gradient of $\hat C$ to the correction network instead of $\hat C$ itself.}
    \label{fig:PINN_diagram3}
\end{figure}

We then move one step further by embedding the corrective forcing term in the dissipation equation directly into the coefficient $\mathcal{C}_{\varepsilon 0}$, while fixing the other coefficients at the final values reported in Tab.~\ref{tab:definitive_coefs}. The main modification, however, is that instead of allowing the correction to depend explicitly on time and on the initial conditions, as before, it is now expressed solely in terms of quantities reconstructed by the model itself; see Fig.~\ref{fig:PINN_diagram3} for a schematic description and  Appendix~\ref{sec:pinncoeff} for details. Despite this loss of information, the resulting PINN model is fully consistent when used autoregressively. Moreover, it still exhibits very good performance after integration and comparison with the test trajectories, as reported in Tab.~\ref{tab:companalytical}.

At this stage, we provide in Tab.~\ref{tab:PINN_details} a summary of the different PINN-based strategies used so far. The final step in deriving the model consists in expressing the neural-network correction for $\mathcal{C}_{\varepsilon 0}$ analytically in terms of a set of suitably chosen quantities that can be computed by the model itself.

Since the objective is to derive a tractable model with minimal variable dependence, we restrict the dataset to inertial trajectories, namely those with $\mathsf{R} \ge 14$ and $\mathsf{D} \le 2$. Seeking an analytical expression for the coefficient over the full dataset would require an additional dependence on the Reynolds number.

The first candidate variable is the Froude number $\Fr$, defined as the ratio of the turbulent frequency to the stratification frequency:
\begin{equation}
    \Fr = \frac{\varepsilon}{k}\frac{1}{\sqrt{2 \partial_z \overline{C}}}.
\end{equation}
This quantity plays a central role in the Rayleigh-Taylor phenomenology, since it characterizes the relative importance of buoyancy effects and turbulent energy transfer across scales.

However, to account for the enstrophy blow-up, one must consider that the energy transfer is not instantaneous and occurs over a finite time scale. As a result, the energy transfer associated with the direct cascade becomes decoupled from dissipation. To mimic this effect, the turbulent kinetic energy can be split into two reservoirs: the directed energy $k_d$, associated with large-scale coherent structures, and the remaining energy, associated with small-scale eddies. Following \cite{Griffond2022}, we define the directed energy as
\begin{equation}
    k_d \coloneqq \frac{1}{2}\frac{\overline{u_z c}^2}{b}
    = \frac{1}{2}\left(\frac{\nu_t}{\sigma_c} \partial_z \overline{C}\right)^2 \frac{1}{b},
    \label{eq}
\end{equation}
where the right-hand side of Eq.~\eqref{eq} follows from the first-gradient closure of the flux in Eq.~\eqref{eq:uc}. From this definition, it follows immediately that $k_d \leq k$. %We will discuss later that this condition may be violated because of the first-gradient closure of the flux.

Therefore, we express $\mathcal{C}_{\varepsilon 0}$ analytically as a function of the Froude number and the directed-to-total kinetic energy ratio, using symbolic regression applied to data generated by the neural-network surrogate. Among the various possible approaches, including Bayesian methods~\cite{Schmelzer2020} or parsimonious regression techniques~\cite{Brunton2016,Thevenin2022}, we adopt a genetic-algorithm-based strategy implemented in the PySR toolbox~\cite{Cranmer2023}, as detailed in Appendix~\ref{app:sym_reg}. This procedure yields the following expression:

\begin{align}
    \Cepsz(\DFr,\DKdK)
          & = 1.252 + \DFr+
    \left[
        \frac{0.113}{(\DFr + 0.555)(\DKdK + 0.397)}
        \right]^{3.185},
    \label{eq:analytical}
    \\
    \DFr  & \coloneqq \Fr - 0.625,
    \notag                          \\
    \DKdK & \coloneqq \KdK - 0.404.
    \notag
\end{align}

Here, the variables are expressed as departures of the Froude number and of the directed-to-total kinetic-energy ratio from their asymptotic values in the self-similar regime. In particular, Eq.~\eqref{eq:analytical} recovers the value $C_{\varepsilon 0}=1.374$ when $\DFr = \DKdK = 0$, which is close to the value obtained from the calibration in the turbulent regime. However, this expression can become singular at low Froude number or low directed-to-total kinetic-energy ratio. Although this does not occur within the dataset, the coefficient $\mathcal{C}_{\varepsilon0}$ is constrained to remain within the interval $[1,4]$.

\begin{figure}[ht!]
    \centering
    \includegraphics[width=0.7 \linewidth]{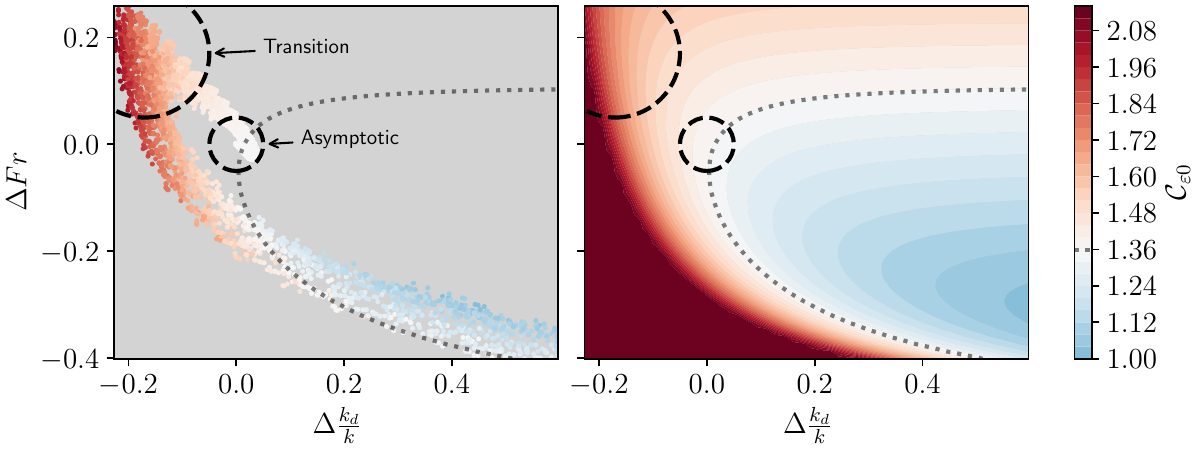}
    \caption{Comparison between the analytical mapping (right) from $\DFr$ and $\DKdK$ to $\Cepsz$ and the true distribution (left) for points sampled on 50 train simulations. The sampling is performed using the methodology detailed in Appendix \ref{app:sym_reg}.}
    \label{fig:sym_reg_mapping}
\end{figure}

In Fig.~\ref{fig:sym_reg_mapping}, we show the evolution of $\mathcal{C}_{\varepsilon 0}$ obtained from the PINN-NC-$\mathcal C_{\varepsilon 0}$ in the $\DFr$--$\DKdK$ plane, together with the analytical correction derived by symbolic regression. The fact that the coefficient remains single-valued in this representation supports our choice of expressing $\mathcal{C}_{\varepsilon 0}$ as a function of these two quantities.
During the transition regime, in the vicinity of the enstrophy blow-up, the coefficient $\mathcal{C}_{\varepsilon 0}$ is therefore slightly increased compared with its original value, allowing the model to reproduce the dissipation peak.

Noticeably, in the new analytical model, the scalar variance $b$ is no longer passive with respect to $k$ and $\varepsilon$, since it directly influences the dissipation equation through the dependence of $\mathcal{C}_{\varepsilon 0}$ on the directed kinetic energy $k_d$.

\begin{figure}[ht!]
    \centering
    \includegraphics[width=1 \linewidth]{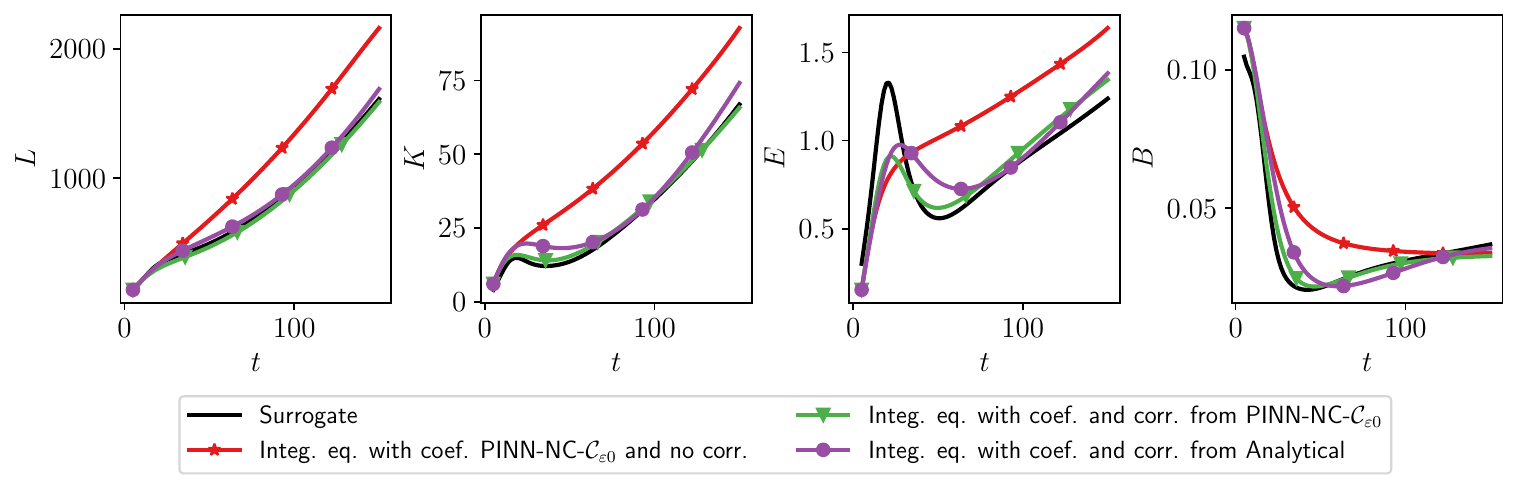}
    \caption{Comparison between the surrogate and three different integrations for a test trajectory (same as figure \ref{fig:foco_integ}). The first integration (red line with stars) is performed with the baseline equations of the model and the coefficients calibrated by the PINN with coefficient correction. The second integration (green line with triangles) includes the correction on $\Cepsz$ predicted with the correction network trained inside the PINN. The third integration (purple line with circles) uses the correction predicted by the analytical model. The reconstruction is less accurate but still captures the peak of dissipation with a correct self-similar growth rate.}
    \label{fig:coco_integ}
\end{figure}

\begin{figure}[ht!]
    \centering
    \includegraphics[width=1 \linewidth]{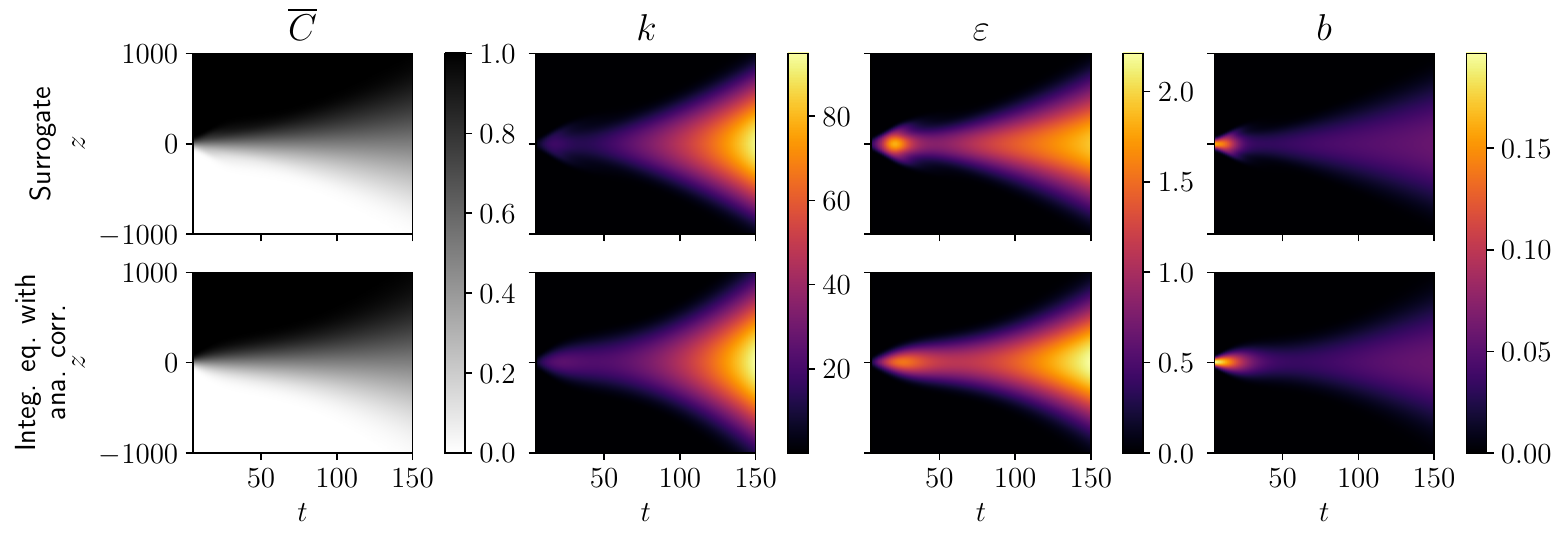}
    \caption{Comparison of the 1D quantities between the surrogate and the analytical model for the usual test simulation. The color scale is the same between the top and bottom graphs. The time axis is composed of 100 points and the integration starts at $t=5$. The vertical axis is composed of 1001 points and $z_\mathrm{max}=1300$. It is cropped here to $z=1000$ for better visibility.}
    \label{fig:coco_integ_1D}
\end{figure}

We now integrate the new analytical model from the profiles provided by the surrogate in order to assess its ability to describe the Rayleigh--Taylor transition. Figure~\ref{fig:coco_integ} presents the evolution of the 0D quantities along a representative trajectory, while Fig.~\ref{fig:coco_integ_1D} shows the corresponding 1D profiles. Although the analytical model performs less well than the PINN-NC-$\mathcal{C}_{\varepsilon 0}$, and is also slightly degraded compared with the fully PINN-C-$f$ model applied to all equations, shown in Fig.~\ref{fig:foco_integ}, its performance remains very satisfactory: it captures the transition reasonably well and ultimately provides a very accurate prediction of the mixing width at final time.

\begin{figure}[ht!]
    \centering
    \begin{subfigure}{1\linewidth}
        \centering
        \includegraphics[width=1 \linewidth]{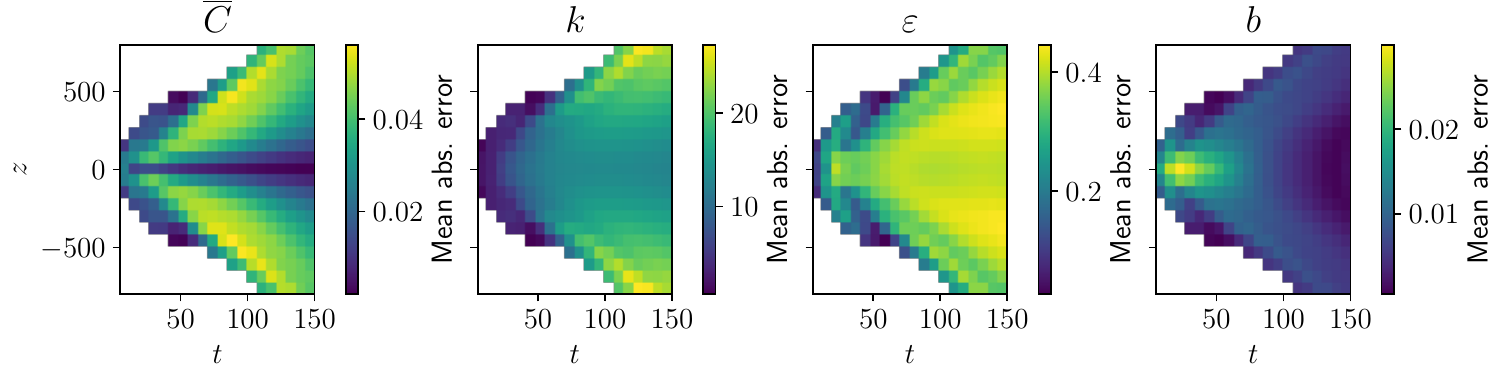}
        \caption{Localization of error for integrated equations without correction}
    \end{subfigure}
    \begin{subfigure}{1\linewidth}
        \centering
        \includegraphics[width=1 \linewidth]{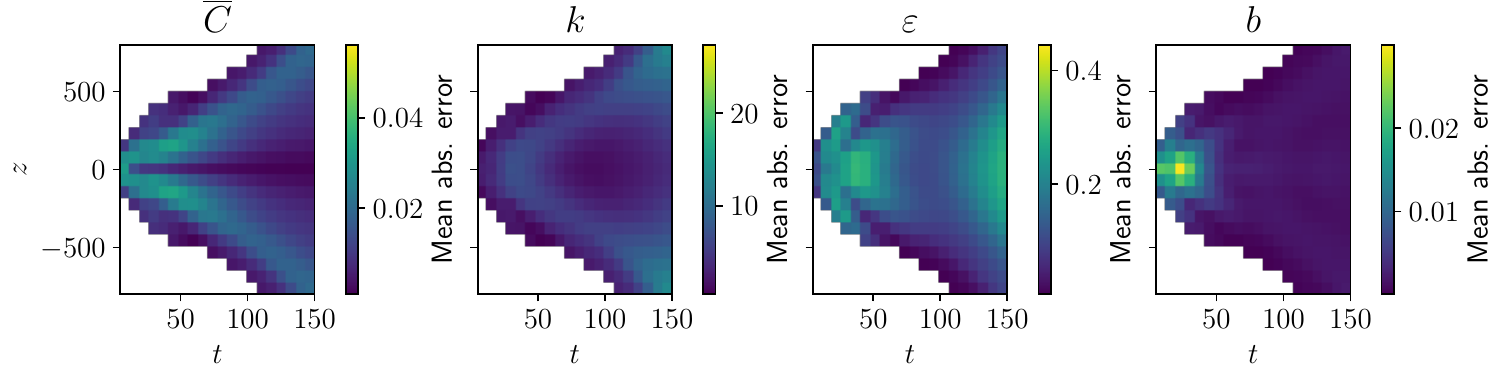}
        \caption{Localization of error for integrated equations with analytical correction}
    \end{subfigure}
    \caption{Comparison of mean absolute error heatmap between integrated equations without and with analytical correction over the entire test set. The error is computed by dividing the time axis into 20 subsets and the vertical axis in 21. Then within each cell, the mean error is computed for all available points. If no points are available, the cell is kept white. The color scale is the same between with and without the correction. Interestingly, the error is very low along the central line of the $\cNN$ profiles because $\cNN$ is always equal to 0.5 at this location.}
    \label{fig:error_heatmap}
\end{figure}

The discrepancy between the reference solution provided by the surrogate model and the baseline or modified $k$-$\varepsilon$-$b$ model can be examined more closely in Fig.~\ref{fig:error_heatmap}. This comparison highlights the improvement brought by the $\mathcal{C}_{\varepsilon 0}$ correction, which boosts dissipation production and enables the model to capture the Rayleigh--Taylor transition to turbulence more accurately than the standard model.

\section{Conclusion}

In this work, we leverage PINNs equipped with various corrective neural networks to calibrate and improve the closure of a $k$-$\varepsilon$-$b$ model for capturing the Rayleigh--Taylor transition to turbulent mixing. Using an extensive dataset of Rayleigh--Taylor DNS parameterized by initial conditions~\cite{Thevenin2025data}, embedded into a high-fidelity surrogate model, we first compare several calibration strategies in the turbulent regime, ranging from simple regression to dynamic Bayesian calibration, in order to assess the ability of the PINN to reproduce these results. After verifying the inability of the $k$-$\varepsilon$-$b$ model to reproduce the transition to turbulence, we use the PINN with corrective terms to disentangle parametric and structural model errors and to identify where physics is missing from the model. As expected, the main discrepancy originates from the dissipation equation, which is not designed to capture the spectral imbalance between the direct energy transfer driven by buoyancy forces and dissipation at small scales. We therefore turn our attention to modifying the dissipation equation, and more specifically its production term, which is regulated by the coefficient $\mathcal{C}_{\varepsilon 0}$. Guided by physical considerations, we seek a dependence of this coefficient on the Froude number, which characterizes the intensity of buoyancy forces relative to turbulent inertia, and on the ratio of directed to total kinetic energy, which quantifies the importance of large-scale coherent structures in the mixing layer. We then apply symbolic regression to the coefficient represented by the corrective neural network in the PINN, thereby deriving an analytical expression that depends on the Froude number and on the directed-to-total kinetic-energy ratio. Notably, through this modification, the scalar variance $b$, which is a passive quantity in the baseline model, becomes active in the dissipation equation. After integration from the initial state, the corrected $k$-$\varepsilon$-$b$ model demonstrates its ability to capture the enstrophy blow-up characterizing the transition to turbulence and to accurately recover the late-time mixing-layer width. Owing to its relatively simple analytical form, the proposed correction can be readily implemented in production codes, improving the representation of the Rayleigh--Taylor transition to turbulence while preserving the robustness of the baseline model.

The methodology based on PINNs with corrective neural networks presented in this work offers a robust strategy for improving any RANS model and can be applied to other types of flows.

\appendix
\section{0D equations of the \texorpdfstring{\KepsS}{K-eps-b} model \label{sec:0D}}
In this section, we propose a 0D formulation of our baseline  $k$-$\varepsilon$-$b$ model. It is derived from the definition of the mixing zone width $L$ in Eq.~\eqref{eq:C}, together with the integration over $z$ of the model equations \eqref{eq:k}, \eqref{eq:e}, and \eqref{eq:s}. Assuming a piecewise profile for the mean concentration, and parabolic profiles for $k$, $\varepsilon$, and $b$ within the mixing zone, we obtain the following system of equations:
\begin{subequations}
    \begin{equation}
        \dot L= 12\frac{\mathcal C_\mu}{\sigma_c} \frac{K^2}{EL},
    \end{equation}
    \begin{equation}
        \dot K=-\frac{\dot L}{L} K +2 \frac{\mathcal C_\mu}{\sigma_c} \frac{K^2}{EL} -E,
    \end{equation}
    \begin{equation}
        \dot E=-\frac{\dot L}{L} E +2 \mathcal C_{\varepsilon 0} \frac{\mathcal C_\mu}{\sigma_c} \frac{K^2}{EL} -\mathcal C_{\varepsilon 2}\frac{E^2}{K},
    \end{equation}
    \begin{equation}
        \dot B=-\frac{\dot L}{L} B +2 \frac{\mathcal C_\mu}{\sigma_c} \frac{K^2}{EL^2} -2 \mathcal C_\chi \frac{E}{K} B.
    \end{equation}
\end{subequations}
The terms on the right-hand side correspond, respectively, to dilution, production, and dissipation. This system facilitates the calibration of the model coefficients, except for those related to diffusion.

\section{Calibration approaches}
In this appendix, we detail the different methods used to calibrate the $k$--$\varepsilon$--$b$ model against data.
\subsection{Linear regression \label{sec:regression}}

To determine the appropriate coefficients, perhaps the most elementary approach is to apply a linear regression to the data using the \KepsS equations. The data are restricted to the time interval $t \in [80,150]$, which corresponds to an \textit{a priori} domain of validity for the model, since the self-similar regime has already begun, although it is not yet fully established.

The equations are rewritten under the form:

\begin{subequations}
    \begin{equation}
        \sigmac = \mathrm{argmin}_{x \in \mathbb{R}} \left\lVert \dot{\LNN} - 12 \frac{\Cmu}{x} \frac{\KNN^2}{\ENN \LNN}\right\rVert^2_2
    \end{equation}
    \begin{equation}
        \Cepsz = \mathrm{argmin}_{x \in \mathbb{R}}  \left\lVert \dot{\ENN} + \frac{\dot{\LNN}}{\LNN} \ENN + \Cepsd \frac{\ENN^2}{\KNN}  - 2 x \frac{\Cmu}{\sigmac}\frac{\KNN}{\LNN} \right\rVert^2_2
    \end{equation}
    \begin{equation}
        \Cchi = \mathrm{argmin}_{x \in \mathbb{R}}  \left\lVert \dot{\VNN} +  \frac{\dot{\LNN}}{\LNN} \VNN - 2 \frac{\Cmu}{\sigmac} \frac{\KNN^2}{\ENN \LNN^2} + 2 x \frac{\ENN}{\KNN} \VNN \right\rVert^2_2
    \end{equation}
\end{subequations}

We note that the equation for $\KNN$ is absent, since it contains no coefficient to calibrate. The regression is first performed for $\sigmac$, and then for $\Cepsz$ and $\Cchi$, which require a prescribed value of $\sigmac$. The time derivatives are computed from the data using finite differences. For each profile in the training set (\ie, $70\%$ of the available initial conditions in the original database), 10 points are sampled at regular intervals.

\subsection{Dynamics Bayesian calibration\label{sec:bayesian}}

This section describes the method used to perform a bayesian calibration of the coefficients using a Monte-Carlo Markov Chains (MCMC) \cite{Metropolis1953} method.

The idea for the calibration is similar to the one performed by Nadiga et al. \cite{Nadiga2019}. The Bayes theorem \cite{Bayes1763} yields:

\begin{equation}
    P(\bm{\beta} | \bm{q}) = \frac{P(\bm{q} | \bm{\beta}) P(\bm{\beta})}{P(\bm{q})}
\end{equation}

with $P(\beta)$ the prior, $P(\bm{\beta} | \bm{q})$ the posterior and $P(\bm{q} | \bm{\beta})$ the likelihood. In practice, $P(\bm{q})$ is unknown and not of great interest as the MCMC algorithm seeks for a function proportional to $P(\bm{\beta} | \bm{q})$, the normalization being performed afterwards to obtain the distribution. It is only required to define a prior and a likelihood model.

We decided to use here an elementary Gaussian model for the likelihood assuming it follows a normal law centered around the surrogate values and independence of the predictions for each quantity \ie $\bm{\Sigma} = \bm{\gamma} \bm{I}$ (the unusual notation $\bm{\gamma}$ for the standard deviation is used to avoid any confusion with the sigmas already appearing in the \KepsS model). This provides an analytical expression for the likelihood

\begin{equation}
    P (\bm{q} | \bm{\beta}) = \frac{1}{(2\pi)^{N_t / 2} \lvert \bm{\Sigma} \rvert^{1/2}} \exp \left[ -\tfrac{1}{2} \left(\pred{\bm{q}} - \bm{q}\right)^T \bm{\Sigma}^{-1} \left(\pred{\bm{q}} - \bm{q}\right) \right]
\end{equation}

In practice, here, $\bm{\beta}$ is the concatenation of the coefficients of the model, \ie for the volume-averaged equations $\sigmac$, $\Cepsz$ and $\Cchi$, and of the standard deviation $\gamma$ for each quantity, $\LNN$, $\KNN$, $\ENN$ and $\VNN$. The vector $\pred{\bm{q}}$ consists in the concatenation of the values of $\pred{q}(t)$ over the time interval of calibration. It is obtained by integrating the \KepsS equations with the coefficients starting from $q(t=t_\mathrm{start})$ as initial conditions.

The prior is chosen to be as little informative as possible, only restraining the values to avoid the divergence of the solver or absurd combinations of parameters. It consists in:
\begin{equation}
    P(\bm{\beta}) = P \begin{pmatrix}
        \sigmac       \\
        \Cepsz        \\
        \Cchi         \\
        \gamma_{\LNN} \\
        \gamma_{\KNN} \\
        \gamma_{\ENN} \\
        \gamma_{\VNN}
    \end{pmatrix}
    \sim \begin{pmatrix}
        \mathcal{U}(0.018,0.45)                  \\
        \mathcal{U}(0.75,2)                      \\
        \mathcal{U}(0.2,5)                       \\
        \mathcal{U}\left(0, \aver{\LNN} \right)  \\
        \mathcal{U}\left(0, \aver{ \KNN} \right) \\
        \mathcal{U}\left(0, \aver{\ENN} \right)  \\
        \mathcal{U}\left(0, \aver{\VNN} \right)  \\
    \end{pmatrix}
\end{equation}

with $\mathcal U$ an uniform distribution defined over an interval and  an upper bound  provided by
\begin{equation}
    \aver{a} =  \frac{1}{30 N_\mathrm{data}} \sum_n a_n.
\end{equation}

It is worth stressing here that, despite the simplicity of the chosen model, it already provides a form of separation between parametric and structural errors. Indeed, after applying the MCMC method, one obtains a probability distribution for each model parameter, \ie, a parametric error, as well as a standard deviation for the predicted quantities. This latter term contains the error that cannot be compensated for by any choice of parameters, \ie, the structural error itself.

\begin{figure}[ht!]
    \centering
    \includegraphics[width=1 \linewidth]{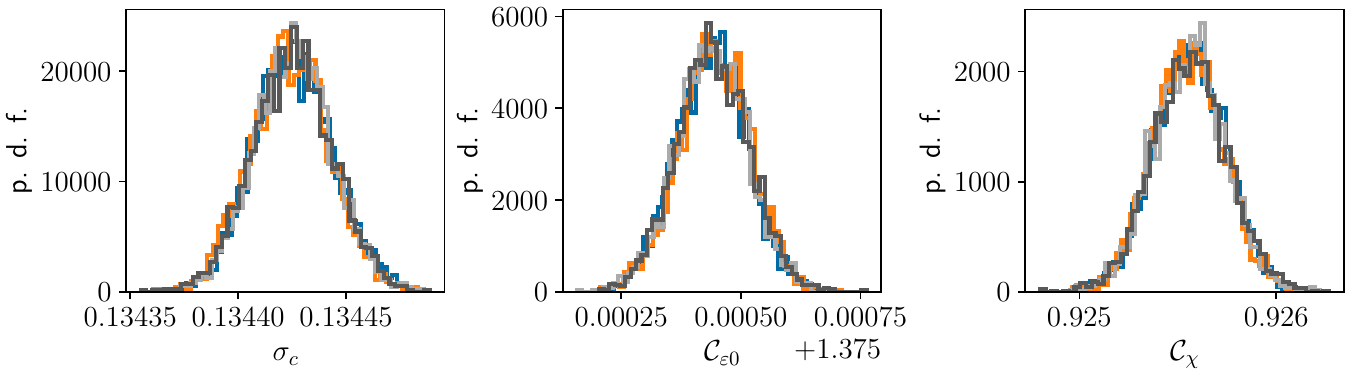}
    \caption{Posterior obtained with the bayesian calibration using the NUTS method for $t\in[80,150]$. Four independent chains are used, each with 3000 burn-in steps and 10 000 sampling steps. The solver used for the integration is RK4 with 240 regularly spaced points in time.}
    \label{fig:mcmc_posterior}
\end{figure}

Finally, the estimation of $P(\bm{\beta} \mid \bm{q})$ is performed using the No-U-Turn Sampler (NUTS) algorithm~\cite{Hoffman2011}, an improved version of Hamiltonian Monte-Carlo (HMC)~\cite{Neal2012} that avoids the need to tune a hyperparameter. The implementation used here is the one included in PyMC~\cite{Pymc2023}. An example of the posterior distributions obtained is shown in Fig.~\ref{fig:mcmc_posterior}.

\subsection{Calibration using physics informed neural-networks, PINN-C \label{sec:pinn}}

This section describes the procedure used to calibrate the coefficients with the PINNs.

\subsubsection{0D and 1D data}

The training is performed using two pools of data. A volume-averaged one that contains $\LNN$, $\KNN$, $\ENN$ and $\VNN$ and depends only on $ t$, \RBSDtext. A plane-averaged pool with $\cNN$, $\kNN$, $\eNN$ and $\vNN$ that has an additional spatial dependency $z$.

The volume-averaged data are sampled using the surrogate model on 100 points regularly spaced between the start time of the calibration (denoted $t_\mathrm{start}$) and the end time ($t_\mathrm{end}$) for each available trajectory in the original database (484 in total). The trajectories are split into train, validation and test subsets composed of respectively 70\%, 10\% and 20\% of the available data.

\begin{figure}[ht!]
    \centering
    \includegraphics[width=1 \linewidth]{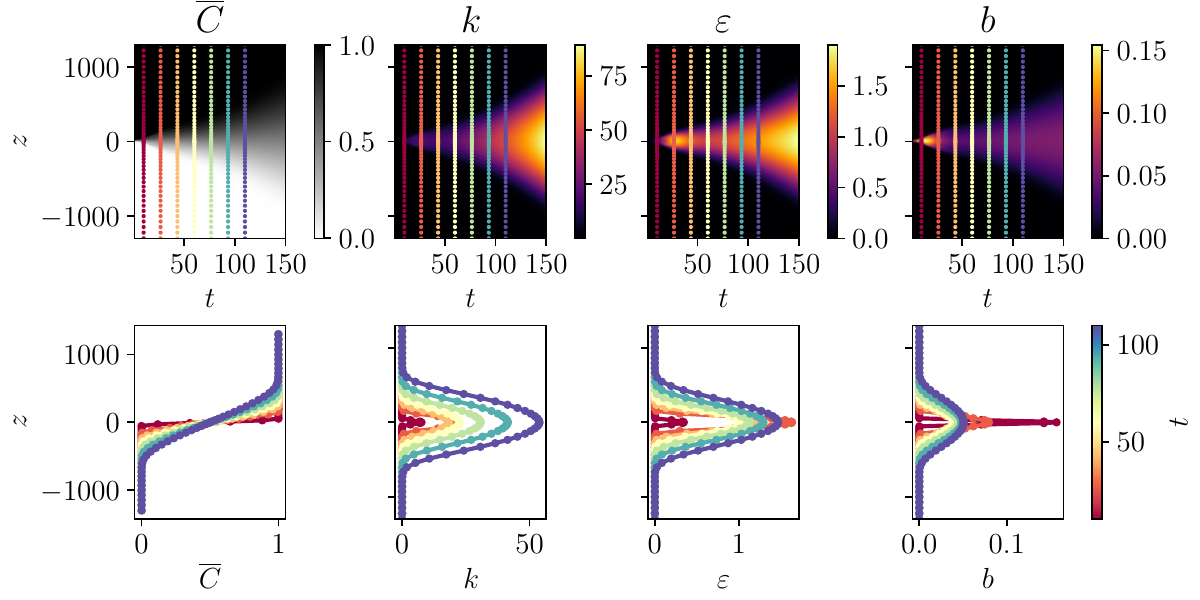}
    \caption{Plane-averaged 1D data used for the training of the PINN-C extracted on a sample trajectory of the dataset. The top row shows the plane-averaged data and each dot correspond to the ones of the profiles. The bottom row corresponds to the shapes of the profiles. Each profile is composed of 51 points sampled with $z_p=1.5$ which increases the concentration of points at the center of the mixing region.}
    \label{fig:1D_data_injection}
\end{figure}

The plane-averaged data consists in profiles, sampled at specific time locations, for 50 random trajectories among the train set. The validation and test sets include the same trajectories as the volume averaged pool. If the calibration is performed for $t\in[1,150]$, seven profiles are included with $t\in \{10,26.6,43.3,60,76.6,93.3,110 \}$ (see Fig.~\ref{fig:1D_data_injection}). For a calibration interval with $t \in [80,150]$, only three profiles are included with $t \in \{85,97.5,110\}$. The goal behind this relatively sparse dataset is to provide the PINN only a partial information on the widths of the profiles of $\cNN$, $\kNN$, $\eNN$ and $\vNN$, and avoid disturbing the training with supplementary information that cannot be captured in the \KepsS framework. Each profile is composed of 51 points sampled using a power uniform distribution with $z_\mathrm{max} = - z_\mathrm{min} = 1300$ and a power of $1.5$. This biased distribution towards the center of the mixing zone was necessary to ensure a sufficient resolution of the profiles even at early times.

\subsubsection{Normalization}

The data normalization is performed using a MinMax method for $t$, $z$, $\RRR$, $\BBB$, $\LNN$, $\KNN$, $\ENN$ and $\VNN$, consisting in

\begin{equation}
    a_{r,n} = \frac{a_n - \min_n(a_n)}{\max_n(a_n) - \min_n(a_n)} \left( a_{r,max} - a_{r,min} \right) + a_{r,min},
\end{equation}

with $a_n$ being the variable of interest and $a_{r,n}$ its normalized version. The parameters $a_{r,max}$ and $a_{r,min}$ are user-defined to choose the bounds of the normalized values. Here respectively 1 and 0 are being used.

The remaining interface parameters $\SSS$ and $\DDD$ are normalized with a LogMinMax:

\begin{equation}
    a_{r,n} = \frac{\log{a_n} - \min_n(\log{a_n})}{\max_n(\log{a_n}) - \min_n(\log{a_n})} \left( a_{r,max} - a_{r,min} \right) + a_{r,min}
\end{equation}

The $\min_n(a_n)$ and $\max_n(a_n)$ are taken over the train dataset to avoid test contamination. For $\LNN$, $\KNN$, $E$ and $\VNN$, $\min(a_n)$ is replaced by 0 to make sure that even the normalized validation and test values remain positive. The plane-averaged variables $\kNN$, $\eNN$ and $\vNN$ are normalized using respectively the same scalers as $\KNN$ and $\ENN$. The concentration $\cNN$ is not normalized as it is already bounded between $0$ and $1$.

As the NN predicts quantities in the normalized space, so does the automatic differentiation. The derivatives are un-normalized to be re-injected in the model's equations using the chain's rule applied to MinMax normalized quantities:

\begin{equation}
    \dd{a_r}{x_r} = \frac{1}{a_{r,max} - a_{r,min}} \frac{\max_n(x_n) - \min_n(x_n)}{\max_n(a_n) - \min_n(a_n)} \dd{a}{x}
\end{equation}

\subsubsection{Colocation domain}

The colocation points are sampled only once before the training and kept fixed during all epochs. They fill a hypercube in the space of $ t$, $ z$, \RBSDtext. The sampling method is Latin Hypercube Sampling \cite{Mckay1979} in which the range of each input variable is first divided into a specified number of equally probable intervals. Samples are then generated by selecting one interval per variable and drawing a value uniformly at random within each interval. The initial sampling is transformed to obtain a power-uniform distribution for $z$ and a log-uniform one for $\mathsf{S}$ and $\mathsf{D}$. Finally, the sampling becomes:

\begin{equation}
    \begin{cases}
        t = d_t                                                                         & \text{with } d_t \sim \mathcal{U}\left(t_\text{start},t_\text{end}\right)                                                                          \\

        z = \frac{d_z}{\left\lvert d_z \right\rvert} \left\lvert d_z \right\rvert^{p_z} & \text{with } d_z \sim \mathcal{U}\left(-z_\text{lim}^{\frac{1}{p_z}}, z_\text{lim}^{\frac{1}{p_z}}\right)                                          \\

        \mathsf{R} = d_\mathsf{R}                                                       & \text{with } d_\mathsf{R} \sim \mathcal{U}\left(0.9 \min_\mathrm{train}(\mathsf{R}),1.1 \max_\mathrm{train}(\mathsf{R})\right)                     \\

        \mathsf{B} = d_\mathsf{B}                                                       & \text{with } d_\mathsf{B} \sim \mathcal{U}\left(0.9 \min_\mathrm{train}(\mathsf{B}),1.1 \max_\mathrm{train}(\mathsf{B})\right)                     \\

        \mathsf{S} = 10^{d_\mathsf{S}}                                                  & \text{with } d_\mathsf{S} \sim \mathcal{U}\left(\log_{10}(0.5 \min_\mathrm{train}(\mathsf{S})),\log_{10}(2 \max_\mathrm{train}(\mathsf{S}))\right) \\

        \mathsf{D} = 10^{d_\mathsf{D}}                                                  & \text{with } d_\mathsf{D} \sim \mathcal{U}\left(\log_{10}(0.5 \min_\mathrm{train}(\mathsf{D})),\log_{10}(2 \max_\mathrm{train}(\mathsf{D}))\right) \\
    \end{cases}
\end{equation}

\subsubsection{Training setup}

The PINN-C takes as input the time $t$, the spatial coordinate $z$, and the parameters of the initial interface perturbation, \RBSDtext. It outputs $\cNN$, $\kNN$, $\omegNN=\eNN/\kNN$, and $\vNN$. Because the values of $\eNN$ in the data are very small, and because $\eNN$ frequently appears in the denominator of the governing equations, training the PINN using the natural output variables $\cNN$, $\kNN$, $\eNN$, and $\vNN$ was initially found to be very tedious. To overcome this difficulty, we reformulated the output variables in a $\kNN$--$\omegNN$ framework and reconstructed $\eNN$ algebraically.

The physics-informed loss term forces the PINN to respect the equations and the coefficients are defined as additional trainable weights that will be tuned during the training. This approach has been used for instance by Shukla et al. \cite{Shukla2020} to determine a velocity field for a sound propagation equation. The supervised loss guides the PINN's dynamic, that then adapts the coefficients to reduce the physics-informed loss. The framework used for the calibration is presented in Fig.~\ref{fig:PINN_diagram}.

The PINN-C used in this work is a fully-connected neural network with 3 hidden layers and 20 neurons per layer. Additional input and output layers are added to plug it on the data. The activation function is Softplus:

\begin{equation}
    \softplus(x) = \frac{1}{\beta} \log \left(1 + \exp(\beta x) \right)
\end{equation}

Its $C^\infty$ property is important here as the model's equations include 2nd order derivatives from the diffusion terms.

The final activation layer is changed to constrain the outputs in a hard way. The functions applied on $\cNN$ and $\kNN$ are respectively a sigmoid and a squared Softplus, and Softplus for $\eNN$. The sigmoid is a hard constraint that forces $\cNN$ and $b$ to remain bounded, while the squared Softplus and Softplus applied on $\kNN$ and $\eNN$ correspond to the expected self-similar time evolution.

An Adam optimiser~\cite{Kingma2014} is used for the first 500 epochs. The training then continues for 10\,000 epochs with the quasi-Newton optimiser SSBroyden~\cite{Urban2025}. This change of optimiser significantly accelerates convergence and allows final loss values to be reached that are far below those obtained with Adam alone. In addition, the collocation points are dynamically weighted using Residual-Based Attention~\cite{Anagnostopoulos2024}. This method relies on an attention vector and emphasizes the collocation points with the largest physics residuals in order to speed up the learning process.

\subsubsection{Loss functions}

The loss function is composed of five terms, $\mathcal{L}_\mathrm{data0D}$ and $\mathcal{L}_\mathrm{data1D}$ to quantify the distance to the volume-averaged and plane-averaged DNS data. $\mathcal{L}_\mathrm{physics}$ to contain the residuals of the model's equations, $\mathcal{L}_\mathrm{boundary}$ that constrains the values at the boundaries of the problem and $\mathcal{L}_\mathrm{coefficient}$ that limits the values taken by the coefficients to avoid a divergence of the training.
The loss function can be written as

\begin{subequations}
    \begin{equation}
        \mathcal{L}_\mathrm{total} = \mathcal{L}_\mathrm{data0D} + \mathcal{L}_\mathrm{data1D} + \mathcal{L}_\mathrm{physics} + \mathcal{L}_\mathrm{coefficient} + \mathcal{L}_\mathrm{boundary}
    \end{equation}

    \begin{equation}
        \mathcal{L}_\mathrm{data1D} = \lambda_\mathrm{data1D} \frac{1}{4 N_\mathrm{data1D}} \sum_n \left\lVert q_{r,n} - \pred{q}_{r,n} \right\rVert^2_2
    \end{equation}

    \begin{equation}
        \begin{cases}
            \pred{\LNN}(t) = \int_{-z_\text{lim}}^{z_\text{lim}} {6 \pred{\cNN}(t,z) \left(1-\pred{\cNN}(t,z) \right) \mathrm{d}z} \\

            \pred{\KNN}(t) = \frac{1}{\pred{\LNN}(t)} \int_{-z_\text{lim}}^{z_\text{lim}} {\pred{\kNN}(t,z) \mathrm{d}z}           \\

            \pred{\ENN}(t) = \frac{1}{\pred{\LNN}(t)} \int_{-z_\text{lim}}^{z_\text{lim}} {\pred{\eNN}(t,z) \mathrm{d}z}           \\

            \pred{\VNN}(t) = \frac{1}{\pred{\LNN}(t)} \int_{-z_\text{lim}}^{z_\text{lim}} {\pred{\vNN}(t,z) \mathrm{d}z}           \\

            \mathcal{L}_\mathrm{data0D} = \lambda_\text{data0D} \frac{1}{4N_\mathrm{data0D}} \sum_n{\left(
                \begin{aligned}
                     & \lambda_\LNN \left( \LNN_{r,n} - \pred{\LNN}_{r,n} \right)^2 \\ + & \left( \KNN_{r,n} - \pred{\KNN}_{r,n} \right)^2 \\ + & \left( \ENN_{r,n} - \pred{\ENN}_{r,n} \right)^2 \\ + & \left( \VNN_{r,n} - \pred{\VNN}_{r,n} \right)^2
                \end{aligned}
                \right)}
        \end{cases}
    \end{equation}

    \begin{equation}
        \begin{cases}
            \mathcal{L}_\cNN = \mathcal{S}_{\dot{\cNN}}^2 \frac{1}{N_\mathrm{coloc}} \sum_n{\left( \partial_t \pred{\cNN}_n -  \partial_z \left[ \left(1 + \frac{\Cmu}{\sigmac} \frac{\pred{\kNN}^2_n}{\pred{\eNN}_n} \right) \partial_z \pred{\cNN}_n \right] \right)^2}                                                                                                                                                                     \\

            \mathcal{L}_\kNN = \mathcal{S}_{\dot{\kNN}}^2 \frac{1}{N_\mathrm{coloc}} \sum_n{\left( \partial_t \pred{\kNN}_n -  \partial_z \left[ \left(1+ \frac{\Cmu}{\sigmaK} \frac{\pred{\kNN}^2_n}{\pred{\eNN}_n} \right) \partial_z \pred{\kNN}_n \right] - 2 \frac{\Cmu}{\sigmac} \frac{\pred{\kNN}^2_n}{\pred{\eNN}_n} \partial_z \pred{\cNN}_n + \pred{\eNN}_n \right)^2}                                                              \\

            \mathcal{L}_\eNN = \mathcal{S}_{\dot{\eNN}}^2 \frac{1}{N_\mathrm{coloc}}\sum_n{ \left( \partial_t \pred{\eNN}_n  - \partial_z \left[ \left(1 + \frac{\Cmu}{\sigmaeps} \frac{\pred{\kNN}^2_n}{\pred{\eNN}_n}\right) \partial_z \pred{\eNN}_n \right] - 2 \Cepsz \frac{\Cmu}{\sigmac} \pred{\kNN}_n \partial_z \pred{\cNN}_n + \Cepsd \frac{\pred{\eNN}^2_n}{\kNN_n} \right)^2}                                                     \\

            \mathcal{L}_\vNN = \mathcal{S}_{\dot{\vNN}}^2 \frac{1}{N_\mathrm{coloc}} \sum_n{ \left( \partial_t \pred{\vNN}_n - \partial_z \left[ \left(1 + \frac{\Cmu}{\sigmaV} \frac{\pred{\kNN}^2_n}{\pred{\eNN}} \right) \partial_z \pred{\vNN}_n \right] - 2 \frac{\Cmu}{\sigmac} \frac{\pred{\kNN}^2_n}{\pred{\eNN}_n} \left( \partial_z \pred{\cNN}_n \right)^2 + 2 \Cchi \frac{\pred{\eNN}_n}{\pred{\kNN}_n} \pred{\vNN}_n \right)^2 } \\

            \mathcal{L}_\mathrm{physics} =  \lambda_\mathrm{physics} \frac{1}{4} \left( \mathcal{L}_c  + \mathcal{L}_\kNN + \mathcal{L}_\eNN + \mathcal{L}_\vNN  \right)
        \end{cases}
    \end{equation}

    \begin{equation}
        \label{eq:coefs_penalty_allD}
        \begin{cases}
            x_0 = -0.12                                                                                                                            \\
            s_{60}(x) = \frac{1}{60} \log{\left( 1 + \exp{\left( 60x \right)} \right)}                                                             \\

            \mathcal{L}_{\sigmac} = s_{60}\left(\frac{\Cmu}{\sigmac} + x_0 - 5\right) + s_{60}\left(0.3 + x_0 - \frac{\Cmu}{\sigmac} \right)       \\

            \mathcal{L}_{\sigmaK} = s_{60}\left(\frac{\Cmu}{\sigmaK} + x_0 - 5\right) + s_{60}\left(0.3 + x_0 - \frac{\Cmu}{\sigmaK} \right)       \\

            \mathcal{L}_{\sigmaeps} = s_{60}\left(\frac{\Cmu}{\sigmaeps} + x_0 - 5\right) + s_{60}\left(0.3 + x_0 - \frac{\Cmu}{\sigmaeps} \right) \\

            \mathcal{L}_{\sigmaV} = s_{60}\left(\frac{\Cmu}{\sigmaV} + x_0 - 5\right) + s_{60}\left(0.3 + x_0 - \frac{\Cmu}{\sigmaV} \right)       \\

            \mathcal{L}_{\Cepsz} = s_{60}\left(\Cepsz + x_0 - 1.92\right) + s_{60}\left(\frac{3}{4} + x_0 - \Cepsz \right)                         \\

            \mathcal{L}_{\Cchi} = s_{60}\left(x_0 - \Cchi \right)                                                                                  \\

            \mathcal{L}_\mathrm{coefficient} = \lambda_\mathrm{coefficient} \left( \mathcal{L}_{\Cepsz} + \mathcal{L}_{\Cchi} + \frac{1}{4} \left( \mathcal{L}_{\sigmac} +  \mathcal{L}_{\sigmaK} + \mathcal{L}_{\sigmaeps} + \mathcal{L}_{\sigmaV} \right) \right)
        \end{cases}
    \end{equation}

    \begin{equation}
        \label{eq:boundary_penalty}
        \mathcal{L}_\mathrm{boundary}  = \lambda_\mathrm{boundary} \frac{1}{4N_\mathrm{boundary}} \sum_n{\left(
            \begin{aligned}
                  & \lambda_{\cNN,\mathrm{bound}} \left( \cNN_{r,n} - \pred{\cNN}_{r,n} \right)^2 \\ + &                 \lambda_{\kNN,\mathrm{bound}} \left( \kNN_{r,n} - \pred{\kNN}_{r,n} \right)^2 \\
                + & \lambda_{\eNN,\mathrm{bound}} \left( \eNN_{r,n} - \pred{\eNN}_{r,n} \right)^2 \\
                + & \lambda_{\vNN,\mathrm{bound}} \left( \vNN_{r,n} - \pred{\vNN}_{r,n} \right)^2
            \end{aligned}
            \right)}
    \end{equation}
\end{subequations}

\begin{figure}[ht!]
    \centering
    \includegraphics[width=1 \linewidth]{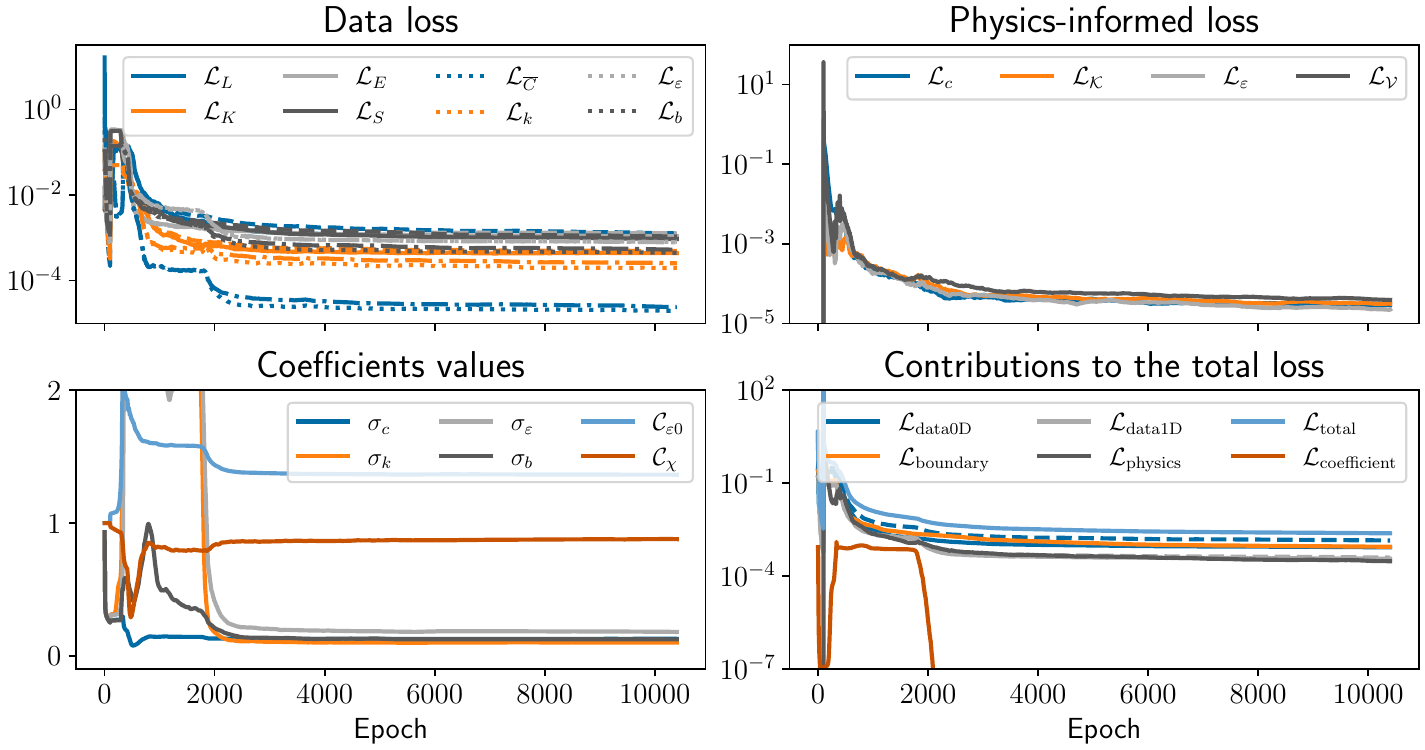}
    \caption{History of training for a PINN calibration with $t_\mathrm{start}=30$. The top left graph corresponds to the supervised terms of the loss, the top right to the physics term, the bottom left to the coefficients values during the training and the bottom right to each weighed term of the loss for which the sum yield $\mathcal{L}_\mathrm{total}$. In top left corner, the validation loss is represented in dashed lines for 0D data and dashed-dotted for 1D.}
    \label{fig:PINN_calibration_history}
\end{figure}

The limits for the coefficients are chosen arbitrarily for the $\sigma$ coefficients, and based on the asymptotic growth rate formula for $\Cepsz$ (see equation \ref{eq:alpha_from_coefs}). The exact values of the hyperparameters listed in the previous formula are given in table \ref{tab:calib_hyperparameters}. An example of training history is presented in Fig.~\ref{fig:PINN_calibration_history}.

\begin{table}[ht!]
    \begin{ruledtabular}
        \begin{tabular}{c|cccccccccc}
            $t_\mathrm{start}$ & $\lambda_\mathrm{data0D}$ & $\lambda_\LNN$ & $\lambda_\mathrm{data1D}$ & $\lambda_\mathrm{boundary}$ & $\lambda_\mathrm{physics}$ & $\lambda_\mathrm{coefficient}$ & $\lambda_{\cNN,\mathrm{bound}}$ & $\lambda_{\kNN,\mathrm{bound}}$ & $\lambda_{\eNN,\mathrm{bound}}$ & $\lambda_{\vNN,\mathrm{bound}}$ \\
            \colrule
            1                  & 1                         & 10             & 1                         & 1                           & 10                         & 0.01                           & 1                               & $10^4$                          & 10                              & 1                               \\
            20                 & 1                         & 10             & 1                         & 1                           & 10                         & 0.01                           & 1                               & 100                             & 10                              & 1                               \\
            50                 & 1                         & 10             & 1                         & 1                           & 10                         & 0.01                           & 1                               & 1                               & 1                               & 1                               \\
            80                 & 1                         & 10             & 1                         & 1                           & 10                         & 0.01                           & 1                               & 1                               & 1                               & 1                               \\
        \end{tabular}
    \end{ruledtabular}
    \caption{Values of the hyperparameters used for the calibration with the PINN. The coefficients for the boundary conditions are modified to ensure they remain well learned even when the initial values are very small.}
    \label{tab:calib_hyperparameters}
\end{table}

\section{Model corrections using physics-informed neural-networks}

\subsection{Force corrections PINN-C-$f$}

This section describes the changes made to the calibration setup to introduce forces in the PINN in order to both calibrate and correct the model.

\subsubsection{Training setup}

The prediction network and the trainable weights associated with the coefficients are left unchanged, and a new neural network is added to predict the correction forces. This network is referred to as the correction network. It takes $(t,z,\RRR,\BBB,\SSS,\DDD)$ as input and returns $(\fococ,\focok,\focoe,\focov)$ as output. For the variant PINN-C-$f_\varepsilon$ with correction forces applied only to the $\eNN$ equation, the output is modified by removing the unused forces. The size and architecture of the correction network are identical to those of the prediction network, with three hidden layers and 20 neurons per layer. However, the final transformation layer is removed, and the activation functions are replaced by hyperbolic tangents.

The renormalized correction forces, denoted with a $r$ subscript, are obtained using the same scaling procedure as for the derivative of their corresponding quantities.

The penalization strategy employed is based on two main ideas. First, the correction forces should be the greatest at the center of the mixing zone and null outside of it. Second, the correction forces at late time are undesirable. To take this into account, the penalization term (detailed in the next section) is a product of the L1 norm of the renormalized forces, of a spatial term and a temporal term. The spatial term expression is $\tfrac{0.25}{\cNN (1-\cNN) + 10^{-3}}$ which is equal to 1 when $\cNN=0.5$ and 250 far from the mixing zone. The temporal term expression is $\left( 1 + \lambda_\mathrm{late} \delta_{t > t_\mathrm{late}}\right)$, which is equal to 1 if $t$ is below the threshold and to $1+\lambda_\mathrm{late}$ otherwise. Typically, $\lambda_\mathrm{late}=100$ and $t_\mathrm{late}=80$. These terms combined allow to guide the correction network towards a physical solution.

An Adam optimiser is used for the first 500 epochs and the training then switches to SSBroyden for a target of 30 000 epochs. The training may be stopped earlier if the Strong Wolfe line search struggles to find acceptable steps.

\subsubsection{Loss functions}

With the introduction of the correction forces, the physics term of the loss $\mathcal{L}_\mathrm{physics}$ is modified to include the correction forces and becomes:

\begin{equation}
    \begin{cases}
        \mathcal{L}_\cNN = \mathcal{S}_{\dot{\cNN}}^2 \frac{1}{N_\mathrm{coloc}} \sum_n{\left( \partial_t \pred{\cNN}_n -  \partial_z \left[ \left(1 + \frac{\Cmu}{\sigmac} \frac{\pred{\kNN}^2_n}{\pred{\eNN}_n} \right) \partial_z \pred{\cNN}_n \right] - \fococ \right)^2}                                                                                                                                                                    \\

        \mathcal{L}_\kNN = \mathcal{S}_{\dot{\kNN}}^2 \frac{1}{N_\mathrm{coloc}} \sum_n{\left( \partial_t \pred{\kNN}_n -  \partial_z \left[ \left(1+ \frac{\Cmu}{\sigmaK} \frac{\pred{\kNN}^2_n}{\pred{\eNN}_n} \right) \partial_z \pred{\kNN}_n \right] - 2 \frac{\Cmu}{\sigmac} \frac{\pred{\kNN}^2_n}{\pred{\eNN}_n} \partial_z \pred{\cNN}_n + \pred{\eNN}_n - \focok \right)^2}                                                             \\

        \mathcal{L}_\eNN = \mathcal{S}_{\dot{\eNN}}^2 \frac{1}{N_\mathrm{coloc}}\sum_n{ \left( \partial_t \pred{\eNN}_n  - \partial_z \left[ \left(1 + \frac{\Cmu}{\sigmaeps} \frac{\pred{\kNN}^2_n}{\pred{\eNN}_n}\right) \partial_z \pred{\eNN}_n \right] - 2 \Cepsz \frac{\Cmu}{\sigmac} \pred{\kNN}_n \partial_z \pred{\cNN}_n + \Cepsd \frac{\pred{\eNN}^2_n}{\kNN_n} - \focoe \right)^2}                                                    \\

        \mathcal{L}_\vNN = \mathcal{S}_{\dot{\vNN}}^2 \frac{1}{N\mathrm{coloc}} \sum_n{ \left( \partial_t \pred{\vNN}_n - \partial_z \left[ \left(1 + \frac{\Cmu}{\sigmaV} \frac{\pred{\kNN}^2_n}{\pred{\eNN}} \right) \partial_z \pred{\vNN}_n \right] - 2 \frac{\Cmu}{\sigmac} \frac{\pred{\kNN}^2_n}{\pred{\eNN}_n} \left( \partial_z \pred{\cNN}_n \right)^2 + 2 \Cchi \frac{\pred{\eNN}_n}{\pred{\kNN}_n} \pred{\vNN}_n - \focov \right)^2 } \\

        \mathcal{L}_\mathrm{physics} =  \lambda_\mathrm{physics} \frac{1}{4} \left( \mathcal{L}_c  + \mathcal{L}_\kNN + \mathcal{L}_\eNN + \mathcal{L}_\vNN  \right)
    \end{cases}
\end{equation}

Furthermore, two new terms are added, one for the penalization of the correction forces $\mathcal{L}_\mathrm{penalty}$, and the other one $\mathcal{L}_{\dot{\LNN} > 0}$ to ensure that despite the correction forces, the growth rate of the mixing length always remain positive. This last term has proven to be useful in a few cases for very early times, improving the quality of the integrated solutions. Their respective expressions are:

\begin{equation}
    \mathcal{L}_\mathrm{penalty} = \lambda_\mathrm{penalty} \left[  \frac{1}{N_\mathrm{coloc}} \sum_n \frac{0.25}{\cNN_n (1-\cNN_n) + 10^{-3}} \left( 1 + \lambda_\mathrm{late} \delta_{t_n > t_\mathrm{late}}\right) \frac{1}{4} \left( \left\lvert \fococ_{r,n} \right\rvert + \left\lvert \focok_{r,n} \right\rvert + \left\lvert \focoe_{r,n} \right\rvert +  \left\lvert \focov_{r,n} \right\rvert \right)  \right]^2
\end{equation}

\begin{equation}
    \mathcal{L}_{\dot{\LNN} > 0} = \frac{1}{N_\mathrm{data0D}} \sum_n \mathrm{ReLu}\left(-\dot{\hat{L}}_{n}\right)
\end{equation}

and so finally, $\mathcal{L}_\mathrm{total}$ becomes:

\begin{equation}
    \mathcal{L}_\mathrm{total} = \mathcal{L}_\mathrm{data0D} + \mathcal{L}_\mathrm{data1D} + \mathcal{L}_\mathrm{physics} + \mathcal{L}_\mathrm{coefficient} + \mathcal{L}_\mathrm{boundary} + \mathcal{L}_{\dot{\LNN} > 0} + \mathcal{L}_\mathrm{penalty}
\end{equation}

\begin{figure}[ht!]
    \centering
    \includegraphics[width=1 \linewidth]{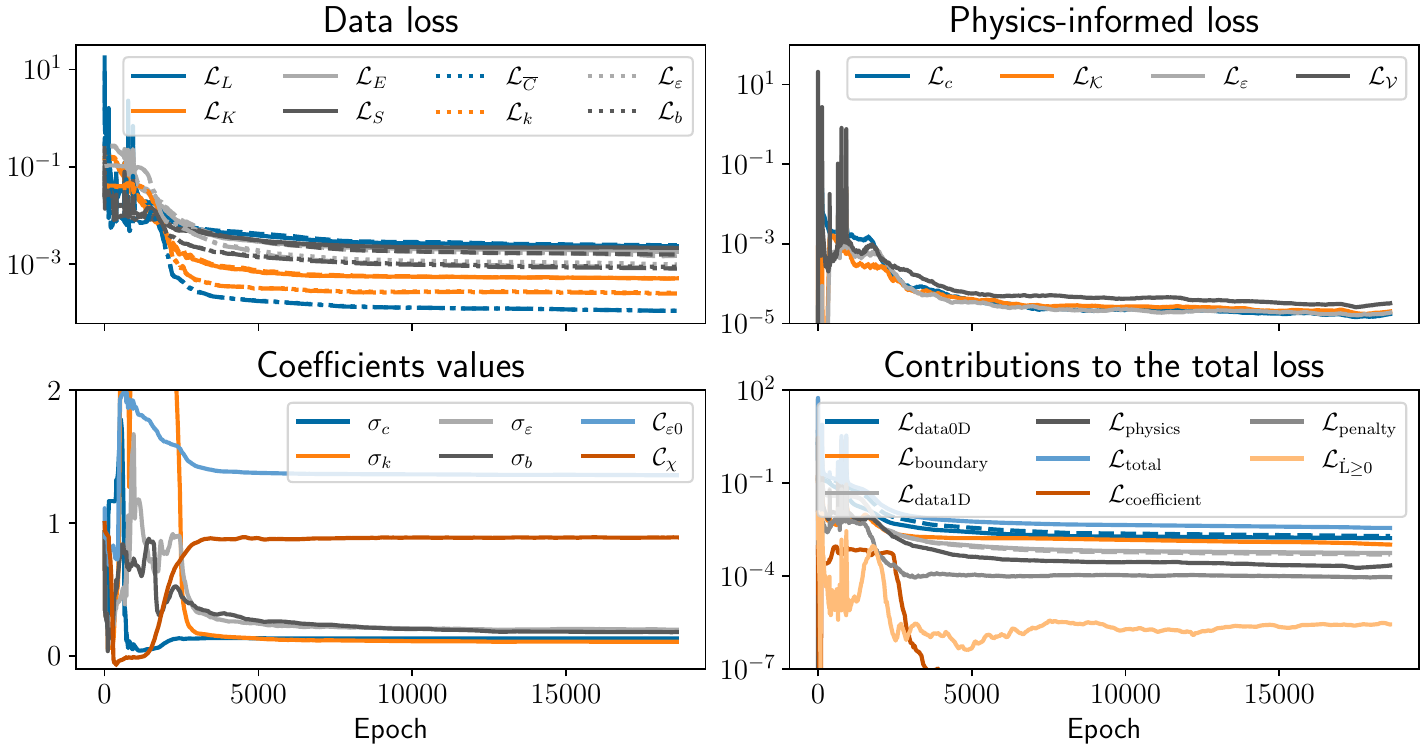}
    \caption{History of training for a PINN with force correction with $\lambda_\mathrm{penalty}=0.001$. Compared to the figure \ref{fig:PINN_calibration_history}, two terms of loss have been added in the lower right graph. One for the penalization of the correction forces and the other to ensure that $\dot{L}$ is always positive.}
    \label{fig:PINN_foco_history}
\end{figure}

An example of the training history obtained for such a model is presented in figure \ref{fig:PINN_foco_history}.

\subsection{Physics informed neural network with coefficient correction PINN-NC-$\mathcal C_{\varepsilon 0}$ \label{sec:pinncoeff}}

This section describes the modifications made to the initial PINN setup to introduce a model correction for the $\Cepsz$ coefficient as presented in Fig.~\ref{fig:PINN_diagram3}.

\subsubsection{Training setup}

As for the force correction, the prediction network is left unchanged. The formerly trainable coefficients are replaced by the constant values reported in Tab.~\ref{tab:definitive_coefs}. These coefficients could in principle have been kept trainable in order to perform calibration and correction simultaneously. This strategy was tested in our experiments, but it made the training more unstable and convergence more difficult, although still manageable. Moreover, it did not provide any significant improvement, since a satisfactory correction was already obtained with the PINN-C-$f_\varepsilon$ model. A new network is added, taking in input $(\partial_z \hat{\cNN}, \hat{\kNN}, \hat{\eNN}, \hat{\vNN})$ and returning in output $\Delta \Cepsz$, the correction part. It has the same number of hidden layers and neurons per layer as the prediction network. The correction network is this time directly plugged onto the outputs of the prediction network. Moreover, an additional automatic differentiation step is required to obtain all the quantities.
The purpose of this method is to leverage the advantages of the PINN by providing a correction that can be computed directly from the state variables of the system. This avoids a tedious projection step, in which an additional model would have to be trained to map the coordinates in simulation space onto the relevant state variables. The resulting correction is therefore fully consistent with the dynamics of the system.

Once again, the correction is penalized through a L1 norm of $\Delta \Cepsz$. No other techniques are necessary to constrain the correction in asymptotic regime or outside the mixing zone.

This framework is more difficult to train because the correction network takes as input the outputs of the prediction network. Therefore, to ensure convergence, it was necessary to adjust the physics-loss weight $\lambda_\mathrm{physics}$. It was changed from a constant into a function of the number of epochs, following a geometric progression: it is initially set to $10^{-6}$ and then increased over 1000 epochs until it reaches its nominal value, denoted here by $\lambda_{\mathrm{physics},\infty}$, at epoch $m_\infty$. Otherwise, the training strategy is similar to that used for the force-correction approach, with 500 epochs of Adam followed by at most 30\,000 epochs with SSBroyden.

\subsubsection{Loss functions}

The physics loss is identical to the one used for calibration except that $\Cepsz$ is no longer a constant and $\lambda_\mathrm{physics}$ evolves with the number of epochs as:

\begin{equation}
    \lambda_\mathrm{physics}(m) =  10^{-6} \left( 10^{6} \lambda_\mathrm{physics,\infty} \right)^{\min(m/m_\infty,1)}
\end{equation}

In practice, similarly to the previous setups, $\lambda_\mathrm{physics,\infty}=10$.

The penalization term $\mathcal{L}_\mathrm{penalty}$ is expressed as:

\begin{equation}
    \mathcal{L}_\mathrm{penalty} = \lambda_\mathrm{penalty} \frac{1}{N_\mathrm{coloc}} \sum_n \left\lvert \Delta \Cepsz \right\rvert
\end{equation}

In practice, an exploration of the impact of $\lambda_\mathrm{penalty}$ has been performed and the final value kept for satisfying correction is $\lambda_\mathrm{penalty}=1$. The method is identical to the one used for force correction. All other hyper-parameters are left unchanged. The term $\mathcal{L}_{\dot{\LNN} > 0}$ is re-used from the force correction framework for similar reasons.

\begin{figure}[ht!]
    \centering
    \includegraphics[width=1 \linewidth]{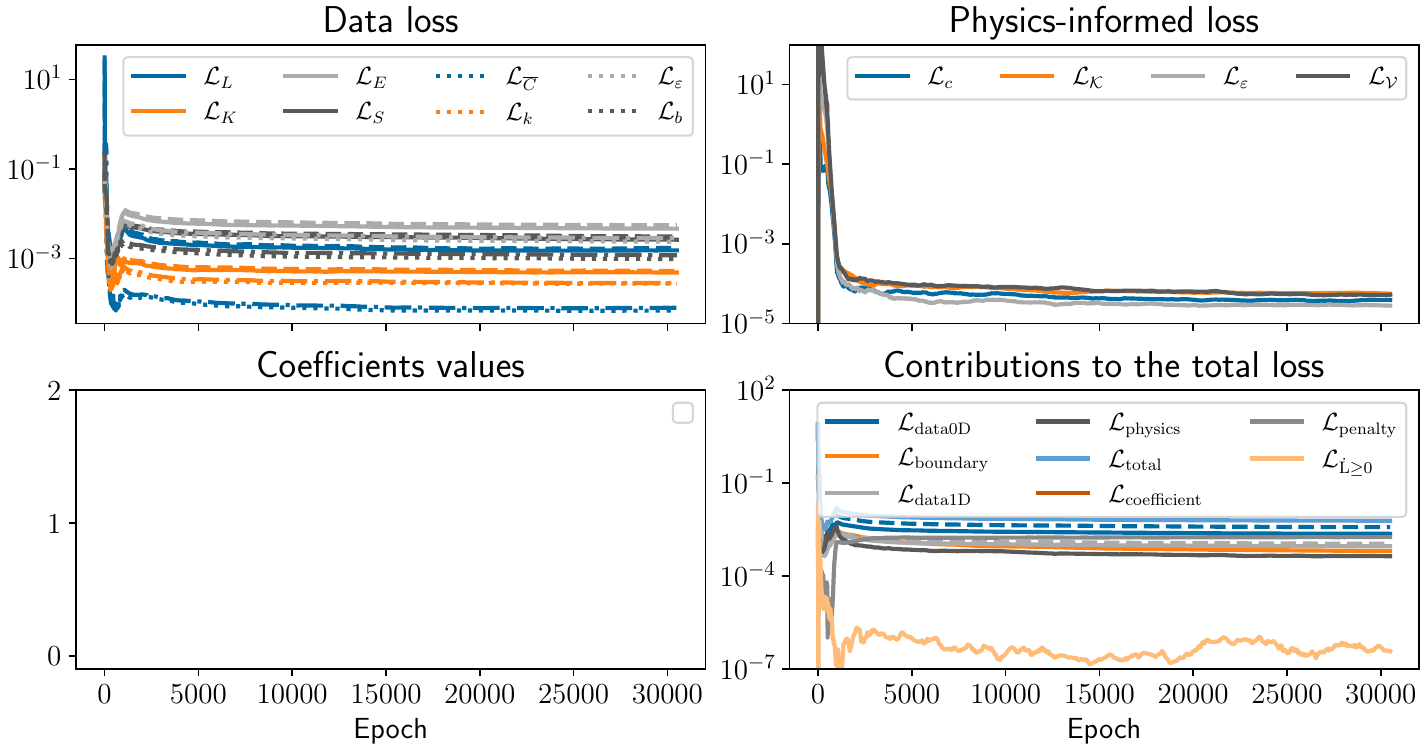}
    \caption{History of the PINN with correction on $\Cepsz$. The lower left plot is empty because no calibration is performed during this training. One can note an increase of the data loss around 1000 epochs, this is due to the increase of $\lambda_\mathrm{physics}$. The prediction network first fits very well the data and progressively degrades the solution to respect the physical constraint as the weight of the physics loss increases.}
    \label{fig:coco_history}
\end{figure}

The training history of the model used in the article is presented in figure \ref{fig:coco_history}.

\section{Symbolic regression for \texorpdfstring{$\Cepsz$}{Cepsz} \label{app:sym_reg}}

The dataset for symbolic regression is composed of two arrays, $X$ and $Y$, the first one containing the values of $\DFr$ and $\DKdK$ obtained by inference of the prediction network and then the formulas. The second one contains the corresponding values of $\Cepsz$ from the correction network. The asymptotic values for the Froude number and the directed to total kinetic energy ratio are computed by averaging the values of $X$ for which $t\geq130$ and $\cNN \in [0.4,0.6]$. The points of the dataset are sampled on 50 train simulations using a power law for 50 points between $\left[ -1300, 1300 \right]$ along the $z$ axis, with $z_p=2$ and another power law with $t_p=1.5$ for 50 points between 5 and 150 along the $t$ axis. Only the points for which $\cNN \in [0.2,0.8]$ are kept to focus on what happens inside the mixing zone where the driving term is significant. This distribution alone was not enough to ensure a stable asymptotic regime. Therefore, 4000 virtual points are inserted in the dataset, chosen randomly in the ball of radius 0.08 centered around $(\DKdK =0.05,\DFr=0)$, with $\Cepsz=1.374$ imposed.

The algorithm used to perform the symbolic regression is the one provided by the Python library PySR~\cite{Cranmer2023}. The binary operators allowed between terms in the equations are addition, multiplication, exponentiation, and division. No unary operators are required. A nested constraint is applied to the exponentiation operator so that expressions can only be raised to constant powers, never to a variable or to another expression. The loss to be minimised is the mean squared error with respect to the ground truth. The final result is obtained after approximately one hour of computation. Since PySR computes the Pareto front of the best expressions, sorted by complexity, the expression that maximises the score, \ie, the negated derivative of the log-loss function with respect to complexity, is retained as the final result.

\section*{Acknowledgements}

DNS and training of the models were performed at the french computing center TGCC.

%------------------------BIBLIOGRAPHIE---------------------------------------
%\bibliographystyle{jfm}
\bibliographystyle{apsrev}
\bibliography{bibprfrt}
\end{document}